\documentclass[epj]{svjour}
\usepackage[utf8]{inputenc}
\usepackage{graphicx}
\usepackage[dvipsnames]{xcolor}
\usepackage[T1]{fontenc}
\usepackage{comment}
\usepackage{amsmath}
\usepackage{widetext}
\usepackage[style=numeric,sorting=none,url=false,doi=false,related=false]{biblatex}

\def\ind#1{{_{\mathrm{#1}}}}
  \newcommand{\vectornorm}[1]{\left|\left|#1\right|\right|}
  \renewcommand{\vec}[1]{\boldsymbol{#1}}
  
  \DeclareFieldInputHandler{title}{}

  \usepackage{soul} %For strikethrough text (\st{})

\title{Regimes of motion of magnetocapillary swimmers}
\author{Alexander Sukhov\inst{1} \and Maxime Hubert\inst{2} \and Galien Grosjean\inst{3\and 4} \and Oleg Trosman\inst{2} \and Sebastian Ziegler\inst{2} \and Ylona Collard\inst{3} \and Nicolas Vandewalle\inst{3} \and Ana-Suncana Smith\inst{2, 5} \and Jens Harting\inst{1, 6}}
\institute{Helmholtz Institute Erlangen-N\"{u}rnberg for Renewable Energy (IEK-11), Forschungszentrum J\"{u}lich, F\"{u}rther Stra{\ss}e 248, 90429 N\"{u}rnberg, Germany \and 
PULS Group, Department of Physics, Interdisciplinary Center for Nanostructured Films, Friedrich-Alexander-Universit\"at  Erlangen-N\"{u}rnberg, Cauerstra{\ss}e 3, 91054 Erlangen, Germany 
\and Universit\'{e} de Li\`{e}ge, GRASP Lab, CESAM Research Unit, All\'{e}e du 6 Ao\^{u}t 19, Li\`{e}ge 4000, Belgium \and IST Austria, Lab Building West, Am Campus 1, 3400 Klosterneuburg, Austria 
\and 
Group for Computational Life Sciences, Division of Physical Chemistry, Ru\dj er Bo\v{s}kovi\'c Institute,
 Bijeni\v{c}ka cesta 54, P.P. 180, 10002 Zagreb,
Croatia
\and
Department of Chemical and Biological Engineering and Department of Physics, Friedrich-Alexander-Universit\"at Erlangen-N\"urnberg, F\"{u}rther Stra{\ss}e 248, 90429 N\"{u}rnberg, Germany 
}
\date{November 2020}
\abstract{
The dynamics of a triangular magnetocapillary swimmer is studied using the lattice Boltzmann method. Performing extensive numerical simulations taking into account the coupled dynamics of the fluid-fluid interface and of magnetic particles floating on it and driven by external magnetic fields we identify several regimes of the swimmer motion. In the regime of high frequencies the swimmer's maximum velocity is centered around the particle's inverse coasting time. Modifying the ratio of surface tension and magnetic forces allows to study the swimmer propagation in the regime of significantly lower frequencies mainly defined by the strength of the magnetocapillary potential. Finally, introducing a constant magnetic contribution in each of the particles in addition to their magnetic moment induced by external fields leads to another regime characterised by strong in-plane swimmer reorientations that resemble experimental observations.
}

\begin{document}
\renewcommand{\mkbibnamefirst}[1]{\textsc{#1}}
\renewcommand{\mkbibnamelast}[1]{\textsc{#1}}
\renewcommand{\mkbibnameprefix}[1]{\textsc{#1}}
\renewcommand{\mkbibnameaffix}[1]{\textsc{#1}}

\maketitle

\section{Introduction}

%\JH{Fix American English vs. British English throughout the paper.} %\AS{Fixed to British one}
Understanding the mechanisms of swimming motion of microorganisms and cells at low Reynolds number is the key to new technologies in biological and medical applications~\cite{Purc77, MeSa17, ElWi15}. Simultaneously with the study of motion of biological objects like bacteria and sperm cells~\cite{FrJu07}, a new class of microscale devices appeared  -- artificial or human-made microswimmers. Many of them are designed in a rather simple way, consisting of a number of interacting microscopic particles powered by external excitations, for instance following the framework of the three-beads swimmer~\cite{NaGo04, GoAj08}. Other examples of artificial microswimmers include magnetically active particles \cite{MaOr15}, Janus particles~\cite{GoSo16}, particles enduring chemo-~\cite{SoGo15}, visco-~\cite{LiMo18}, gravi-~\cite{RuSt20} or thermo-taxis~\cite{YaWy14} or even swarms of microscopic particles mimicking the behaviour of biological organisms~\cite{AhBa17}.       

A particular example of an artificial microswimmer capable of self-propelling at a gas/liquid interface is a magnetocapillary microswimmer. Here, several magnetic particles are placed onto an air/water interface. Their assembly is achieved via balancing attractive capillary and repulsive magnetic interactions~\cite{VaCl12}. The motion is induced by applying periodically altered magnetic fields and it can self-propel in a linear~\cite{GrHu16} or a triangular configuration~\cite{LuOb13, GrLa15} or perform fully controlled rotations at the interface~\cite{GrHu19} offering a number of potential applications. These include the transport of cargo particles or interfacial mixing~\cite{GrHu17}. 

Although a number of theoretical studies are known for the triangular swimmer configuration~\cite{RiFa18,RiFa18a,ZiHu19}, many of them disregard the presence of the interface or consider external forces only effectively. Here, we study numerically the rich dynamics of magnetocapillary swimmers by taking all relevant effects into account. In ref.~\cite{SuZi19} we thoroughly investigated the assembly and the motion of the magnetocapillary swimmer in the regime where the peak velocities of motion are centred at frequencies around the inverse viscous time of a single particle. This regime appears quite different from what is observed in the experiments~\cite{GrLa15, GrHu17}: 
1) we do not observe sizable in-plane rotations of the beads and of the swimmer, 2) the translational amplitudes of the bead motion are significantly smaller, 3) the simulated average velocity of the swimmer is lower than in the experiment.
Additionally, the particles detach from the interface and sink at low excitation frequencies, limiting our study to high-frequency regimes.

The current paper aims at a thorough understanding of the parameters that determine the collective motion of the swimmer beads and the propagation efficiency of the full magnetocapillary swimmer. 
We demonstrate and explain the various modes of motion magnetocapillary swimmers can undergo depending on the precise setup and choice of parameters. Finally, we demonstrate that our simulations are also able to qualitatively reproduce the strong reorientations of the swimmer as observed in the experiments.

To do so, we reconsider some of the assumptions made in the previous numerical model. For example, in order to prevent sinking, the ratio of the surface tension and the magnetic forces needs to be strongly modified. Furthermore, the assumption of purely paramagnetic moments in each bead, \textit{i.e.} induced by external magnetic fields only, is not sufficient to describe the regime observed in the experiments. A constant magnetic contribution in each of the particles leads to the experimentally observed in-plane rotations of the particles. 

The remainder of this article is organised as follows: sect. 2 deals with the details of the numerical method, in sect. 3 different regimes of motion are presented and analyzed in depth. Main conclusions on the present and our previous numerical simulations are summarized in the final section.  

\begin{comment}
\begin{figure}[htb]
\centering
\includegraphics[width=0.48\textwidth]{Fig_1.png}
\caption{Schematics of different regimes of motion.}
\label{fig_1}
\end{figure} 
\end{comment}

\section{Simulation method}
The simulation method is thoroughly described in ref.~\cite{SuZi19} and we only summarize the main ingredients here. We use a lattice  Boltzmann (LB) method for the simulation of fluids~\cite{Benzi1992}. It is based on a discretised version of the Boltzmann equation
\begin{equation}
    \displaystyle f_i^{c}(\vec{x}+\vec{c}_i\Delta t, t+\Delta t) = f^{c}_i(\vec{x},t)+\Omega^{c}_i(\vec{x},t). 
\label{eq:LBE}    
\end{equation}
The latter describes the time evolution of a single-particle distribution function $f^{c}_i(\vec{x},t)$ at time $t$ and position $\vec{x}$ and $\vec{c}_i$ denotes the discrete velocity vector in the $i$th direction for fluid component $c=\{1, 2\}$. Here, we use a so-called D3Q19 lattice with $i=1, \ldots, 19$~\cite{Qian1992}. The left hand side of eq.~(\ref{eq:LBE}) describes the free streaming of fluid particles, while their collisions are modelled by a Bhatnagar-Gross-Krook (BGK) collision operator on the right hand side as~\cite{BhGr54}
\begin{equation}
    \displaystyle \Omega^{c}_i(\vec{x},t) = - \frac{f^{c}_i(\vec{x},t)-f^{\mathrm{eq}}_i(\rho^{c}(\vec{x},t),\vec{u}^{c}(\vec{x},t))}{\tau^{c}/\Delta t}. 
\label{eq:BGK}
\end{equation}
In eq.~(\ref{eq:BGK}), $f^{\mathrm{eq}}_i(\rho^{c}(\vec{x},t),\vec{u}^{c}(\vec{x},t))$ is a third-order equilibrium distribution function~\cite{HVC04}, and macroscopic densities and velocities are given by $\rho^{c}(\vec{x},t)=\rho_0\sum_i f^{c}_i(\vec{x},t)$ as well as $\vec{u}^c(\vec{x},t)=\sum_i f^c_i(\vec{x},t)\vec{c}_i/\rho^c(\vec{x},t)$, respectively ($\rho_0$ is a reference density). $\tau\ind{c}$ is the relaxation rate of component $c$, which determines the relaxation of $f^{c}_i(\vec{x},t)$ towards the equilibrium. Space is discretised on a three-dimensional lattice with lattice constant $\Delta x$ and the time $t$ is discretised with $\Delta t$-steps. The speed of sound $c\ind{s}=1/\sqrt{3}\Delta x / \Delta t$ depends on the choice of the lattice geometry and allows one to obtain the kinematic $\nu^{c}=c^2_{\mathrm{s}}\Delta t (\tau^{c}/\Delta t - 1/2)$ or the dynamic $\eta^{c}=\nu^{c}\rho^{c}$ fluid viscosities. For simplicity, we set $\Delta x=\Delta t=\rho_0=\tau^{c}=1$ in the remainder of this paper and refer to the units as lattice units (l.u.).

For simulations of the interface and the associated capillary interactions, we choose the pseudopotential method of Shan and Chen and apply a mean-field force between different fluid components as~\cite{ShCh93,Liu2016} 
\begin{equation}
    \displaystyle \vec{F}^c_{\mathrm{C}}(\vec{x},t) = - \psi^c(\vec{x},t)\sum_{c'}g_{cc'}\sum_{\vec{x}'}\psi^{c'}(\vec{x}',t)(\vec{x}'-\vec{x}).
\label{eq:SC}
\end{equation}
Here, $c$ and $c'$ refer to different fluid components, $\vec{x}'$ denotes the nearest neighbours of the lattice site $\vec{x}$ and $g_{cc'}$ describes a coupling constant determining the surface tension. $\psi^c(\vec{x},t)$ has the form $\psi^c(\vec{x},t)\equiv \psi^c(\rho^c(\vec{x},t))=1-\mathrm{e}^{-\rho^c(\vec{x},t)}$. The force (\ref{eq:SC}) is applied to the fluid component $c$ by adding a shift $\Delta \vec{u}^c(\vec{x},t)=\tau^c\vec{F}^c_{\mathrm{C}}(\vec{x},t)/\rho^c(\vec{x},t)$ to the velocity $\vec{u}^c(\vec{x},t)$ in the equilibrium distribution. The method is a diffuse interface method, with an interface width of typically $5$ lattice sites depending weakly on the coupling strength~\cite{FrGu12, KrFr13}. In the binary fluid system we refer to the fluids as ``red'' (r) and ``blue'' (b)~\cite{JaHa11}. 
%\MH{With respect to the experiments, which one is the air and the water? What are the densities and viscosities?} \JH{Not as simple: both fluids are identical in the method and thus very different from a real air/water system. We on purpose did not make a QUANTITATIVE link to the experiment.}
In addition, we initialize the system with two equally sized volumes of red and blue fluid, separated by a flat fluid interface. 
%\MH{Don't understand. Why not saying that the simulation box is separated in 2 identical volumes of each fluids?
%JH: changed text to "volumes" 
%}

Three rigid magnetic particles are simulated by solving Newton's equations of motion for translational and rotational degrees of freedom by means of a leap-frog algorithm. The particles are discretised on the lattice. They are coupled to both fluid species by means of a modified bounce-back boundary condition for both fluid components~\cite{Ladd94,LaVe01,KHH04,JaHa11,GFH14}. 

\begin{figure}[htb]
\centering
\includegraphics[width=0.48\textwidth]{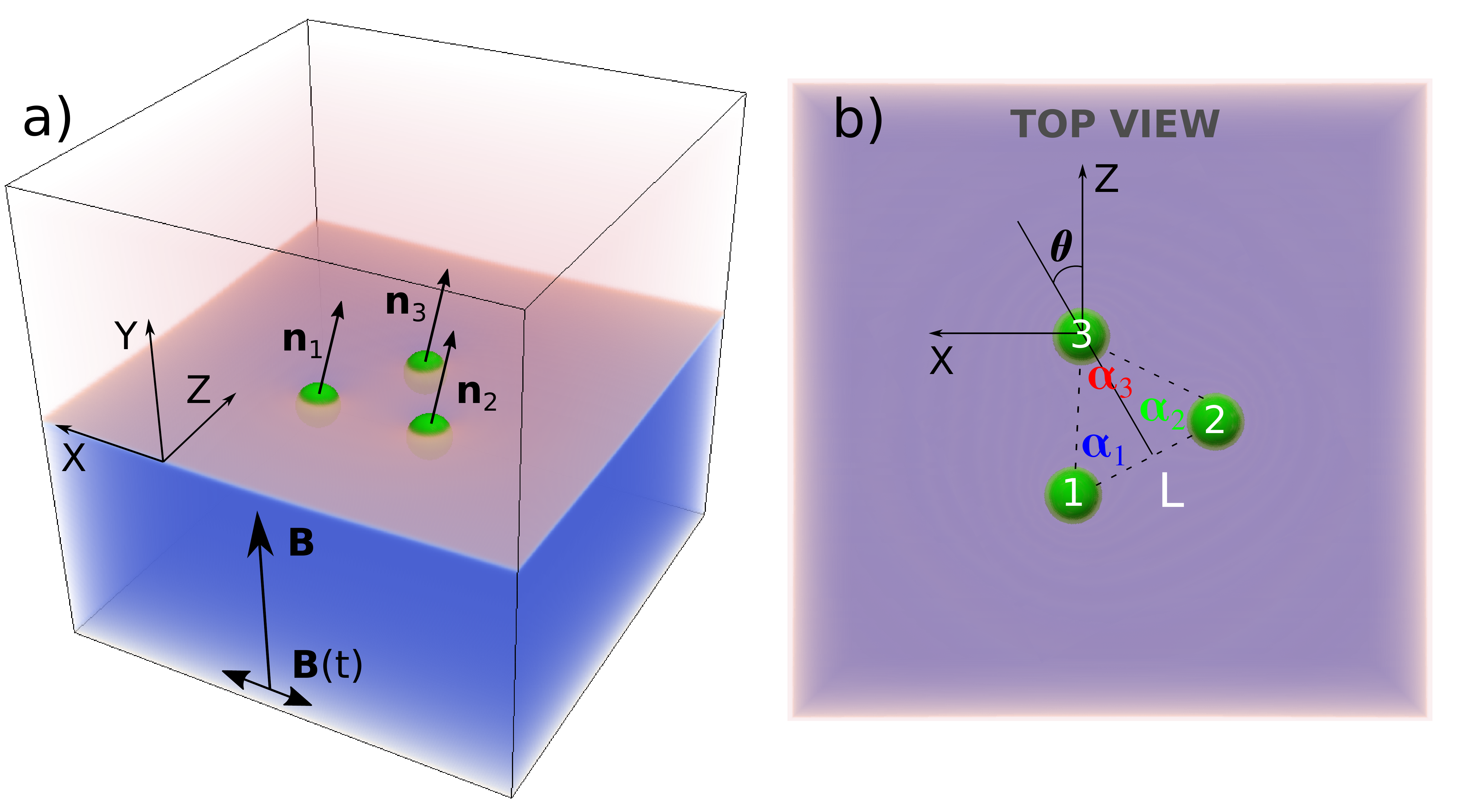}
\caption{a) The simulated system including the directions of external magnetic fields ($\vec{B}$ and $\vec{B}(t)$) and particle orientation vectors $\vec{n}_i$, 
%\AS{
coinciding with the directions of magnetic moments of beads $\vec{\mu}_i$.
%} \MH{couldn't find a definition of ni before. It's really needed (see later)}. 
b) Top view of the fluid interface showing angles $\alpha_i$ of the isosceles triangle formed by particles and the orientation of the triangle within the interface $\theta$.}
\label{fig_1}
\end{figure}

A static magnetic field $B_{\mathrm{y}}$ is applied along the positive $y$-direction (see Fig.~\ref{fig_1}) perpendicular to the interface and induces repulsive magnetic dipolar forces. The repulsion is balanced by an attractive capillary force which is due to the interface deformation caused by the gravity-induced immersion of the particles. This combination of forces allows the assembly of stable particle arrangements at the interface. In analogy with the experiments on magnetocapillary swimmers~\cite{LuOb13, GrLa15} we choose the amplitude of the time-dependent magnetic field to be approximately three times lower than that of the static field to treat it as a modulation. The field $\vec{B}(t)=B_{0 \mathrm{x}}\cos \omega t \vec{e}_{\mathrm{x}}$ 
causes a deformation of the particle arrangement, which due to collective hydrodynamic interactions leads to the motion of the swimmer under a force free protocol.
%
%\MH{Be extra careful here. B(t) is not a force leading to the collective motion. 
%Indeed, once might think that there is a magentic gradient somewhere. I'd 
%state that the field B(t) trigger the deformation of the structure which, by cooperative hydrodynamics effect, leads to the collective motion of the beads under a force free protocol.} \JH{I changed the text. OK?}
%
To describe the paramagnetic nature of the particles, a homogeneous external magnetic field $\vec{B}$ induces a magnetic moment $\vec{\mu}_i=\chi V \vec{B}/\mu_0$ in each particle $i$, where $\chi$ is the particle susceptibility, $V$ is its volume and $\mu_0 = 4\pi \times 10^{-7}$ (in lattice units) corresponds to the magnetic permeability of vacuum. The resulting magnetic dipole-dipole interaction between a pair of particles is 
\begin{equation}
 \displaystyle U_{ij} = -\frac{\mu_0}{4\pi r_{ij}^3}\left[ 3(\vec{\mu}_i\cdot \vec{e}_{ij})(\vec{\mu}_j\cdot \vec{e}_{ij})-(\vec{\mu}_i\cdot \vec{\mu}_j)\right].
\label{eq:Umagn}
\end{equation}
In eq.~(\ref{eq:Umagn}), $r_{ij} \equiv \vectornorm{\vec{r}_{ij}} \equiv \vectornorm{ \vec{r}_i - \vec{r}_j}$ is the distance between the centres of two spheres $i,j$ located at $\vec{r}_i$ and $\vec{r}_j$, respectively, and $\vec{e}_{ij}$ = $(\vec{r}_{i}-\vec{r}_{j})/\vectornorm{\vec{r}_{i}-\vec{r}_{j}}$. 
The effective magnetic field generated by the magnetic moment $\vec{\mu}_j$ at the location of another particle $i$ is
\begin{equation}
 \displaystyle  \vec{B}_i= - \frac{\partial U_{ij}}{\partial \vec{\mu}_i} = \frac{\mu_0}{4\pi r^3_{ij}}\left[3\vec{e}_{ji}(\vec{\mu}_j\cdot \vec{e}_{ji})-\vec{\mu}_j\right].
\label{eq:Bmagn}
\end{equation}
The resulting magnetic force acting on the $i$th particle is then $\vec{F}_i= - \vec{\nabla} \left(- \vec{\mu}_i\cdot (\vec{B}_i+\vec{B})\right)$, or more explicitly 
%\MH{Once again, make sure no one think a grad of magnetic field drive the structure in a given direction}
\begin{eqnarray}
\displaystyle 
 \vec{F}_{i} =  \frac{3\mu_0}{4\pi r^4_{ij}}  && 
\left( \vec{\mu}_i \left( \vec{\mu}_j \cdot \vec{e}_{ji} \right) 
+ \vec{\mu}_j \left(\vec{\mu}_i \cdot \vec{e}_{ji} \right) \right.  \nonumber \\
&& \left. - 5 \vec{e}_{ji} \left(\vec{\mu}_j \cdot \vec{e}_{ji} \right) \left(\vec{\mu}_i \cdot \vec{e}_{ji} \right) +\vec{e}_{ji}(\vec{\mu}_i \cdot \vec{\mu}_j)\right).
\label{eq:Fmagn}
\end{eqnarray}
%\AS{
We note that the external magnetic field $\vec{B}$ is homogeneous ($\vec{\nabla}(\vec{\mu}_i\cdot \vec{B})=0)$), hence the magnetic forces (eq. (\ref{eq:Fmagn})) appear solely as a result of the magnetic dipolar interaction.
%} 
Analogously, the magnetic torque acting on the particle $i$ is $\vec{T}_i=\left[\vec{\mu}_i \times (\vec{B}_i+\vec{B})\right]$, or explicitly 
\begin{equation}
\vec{T}_{i}\!\! = \!\!\frac{\mu_0}{4\pi r_{ij}^3}\cdot \left(3\left(\vec{\mu}_j \cdot \vec{e}_{ji}\right)\left[\vec{\mu}_j\!\times\! \vec{e}_{ji} \right] - \left[\vec{\mu}_i \!\times\! \vec{\mu}_j\right]\right) + \left[\vec{\mu}_i\! \times\! \vec{B}\right].\!\!\!
\label{eq:Tmagn}
\end{equation}
In the case of three particles the total force and the total torque for each particle include a summation of expressions (\ref{eq:Fmagn}) and (\ref{eq:Tmagn}) over index $j$. The method with implemented magnetic interactions has already been benchmarked and successfully applied for simulations of magnetocapillary phenomena \cite{XiDa15, XiDa16, XiDa17} and swimmers \cite{SuZi19}. 

The following numerical parameters are used throughout the paper: the simulation box consists of $128^3$ cubic cells containing two equally sized fluid lamellae. Rigid walls with midgrid bounce back boundary conditions are placed parallel to the fluid interface, while in any other directions periodic boundary conditions are assumed. All beads have equal radius $R=5\Delta x$ and density $\rho\ind{p}=2\rho_0$. The coupling constant $g_{cc'}=0.1$ between the two fluids with densities $\rho_r=\rho_b=0.7\rho_0$ implies a numerical surface tension $\gamma=0.04$ in lattice units. The magnetic moment is chosen in the range $\mu=[1;3]\times 10^5$ in lattice units.

\section{Results}
The equilibrium properties of one, two and three particles at the fluid-fluid interface, are thoroughly studied in ref.~\cite{SuZi19}. Therefore, here we start directly with three rigid magnetic particles placed at the interface. The particles are in their equilibrium position at a fixed ratio of gravitational and surface tension forces, termed as Bond number, \textit{i.e.} $Bo=0.16$ (Fig.~\ref{fig_1}a). 

The assembly of the three particles is driven by a time-dependent magnetic field and the main observable of interest is the average velocity of the swimmer, defined as 
\begin{equation}
\displaystyle \left< \vec{v} \right>=\frac{1}{3}\sum_i \frac{\left(\vec{r}_i(t_{\mathrm{e}})-\vec{r}_i(t_{\mathrm{b}})\right)}{(t_{\mathrm{e}}-t_{\mathrm{b}})}, 
\label{eq_v_swim}
\end{equation}
where $t_{\mathrm{b}}$, $t_{\mathrm{e}}$ stand for the beginning and end times of the external magnetic field action, $i$ numbers the particles and $\vec{r}_i$ denotes the corresponding coordinates. 

%\AS{
In general, the velocities and times can be expressed in relative units related to characteristic processes of the particles at the interface. Since each spherical particle in a fluid experiences a drag force upon translation, its characteristic time to reach the equilibrium can be measured via the coasting or viscous time, defined as $\tau_{\mathrm{cs}}=m/(6\pi \eta R)$ or $\tau_{\mathrm{cs}}=2\rho_{\mathrm{p}}R^2/(9\eta)\approx 95$~$\Delta t$, where $\rho_{\mathrm{p}}$ is the particle density, $R$ is its radius and $\eta$ is the total fluid viscosity.
Following the total magnetic field, the particles partly rotate in the fluid, requiring another relevant time scale associated with their rotation. This rotational time can be defined as a ratio of the moment of inertia and the mechanical torque, \textit{i.e.} $\tau_{\mathrm{rt}}=2/5mR^2/(8\pi \eta R^3)$ or $\tau_{\mathrm{rt}}=\rho_{\mathrm{p}}R^2/(15\eta)\approx 29$~$\Delta t$. One can estimate the time describing the relaxation of the interface as $\tau_{\mathrm{in}}=\eta R/\gamma \approx 44$~$\Delta t$, where $\gamma$ describes the surface tension. Finally, since the particles are in the magnetocapillary potential, the time related to its strength or effective spring $k$ can be approximated using $\tau_{\mathrm{sp}}=2\pi\sqrt{m/k}\approx 65000$~$\Delta t$. It is mostly convenient to use one of the shortest time scales associated with the colloid for the normalization and the time having fewest variable parameters. Therefore, we express time in units of the coasting time of a single particle $\tau\ind{cs}$ and the swimmer velocities in diameters per coasting time $(2R)/\tau_{\mathrm{cs}}$.
%}   \MH{Give the values of all time scales. Also, nothing is compared to those time scale (to the exception of T viscous) which make further discussion difficult to understand/follow. }

Following the definition of the coasting time $\tau\ind{cs}$, we denote \textit{high frequencies} to be in the range of $\omega/(2\pi) \approx 1/\tau\ind{cs}$ 
%\MH{Personal opinion: I wouldn't define high frequency as $\omega/(2\pi) \approx 1/\tau\ind{cs}$}\JH{I agree that it sounds a bit odd, but one has to find a ay to distinguish...}
, while the range of \textit{low frequencies} corresponds to $\omega/(2\pi) \ll 1/\tau\ind{cs}$.
%
%\MH{For me, above can still be stored into the methods section}

\subsection{Motion at high frequencies}
In ref.~\cite{SuZi19} we 
report on the static and some dynamic properties of the magnetocapillary swimmer. It is shown there that the swimmer demonstrates a stable controlled motion for a broad range of swimmer sizes at frequencies in the vicinity of the inverse coasting time $\tau_{\mathrm{cs}}$. %\MH{Don't understand this %paragraph. JH: Changed %wording.}

\begin{figure}[htb]
\centering
\includegraphics[width=0.48\textwidth]{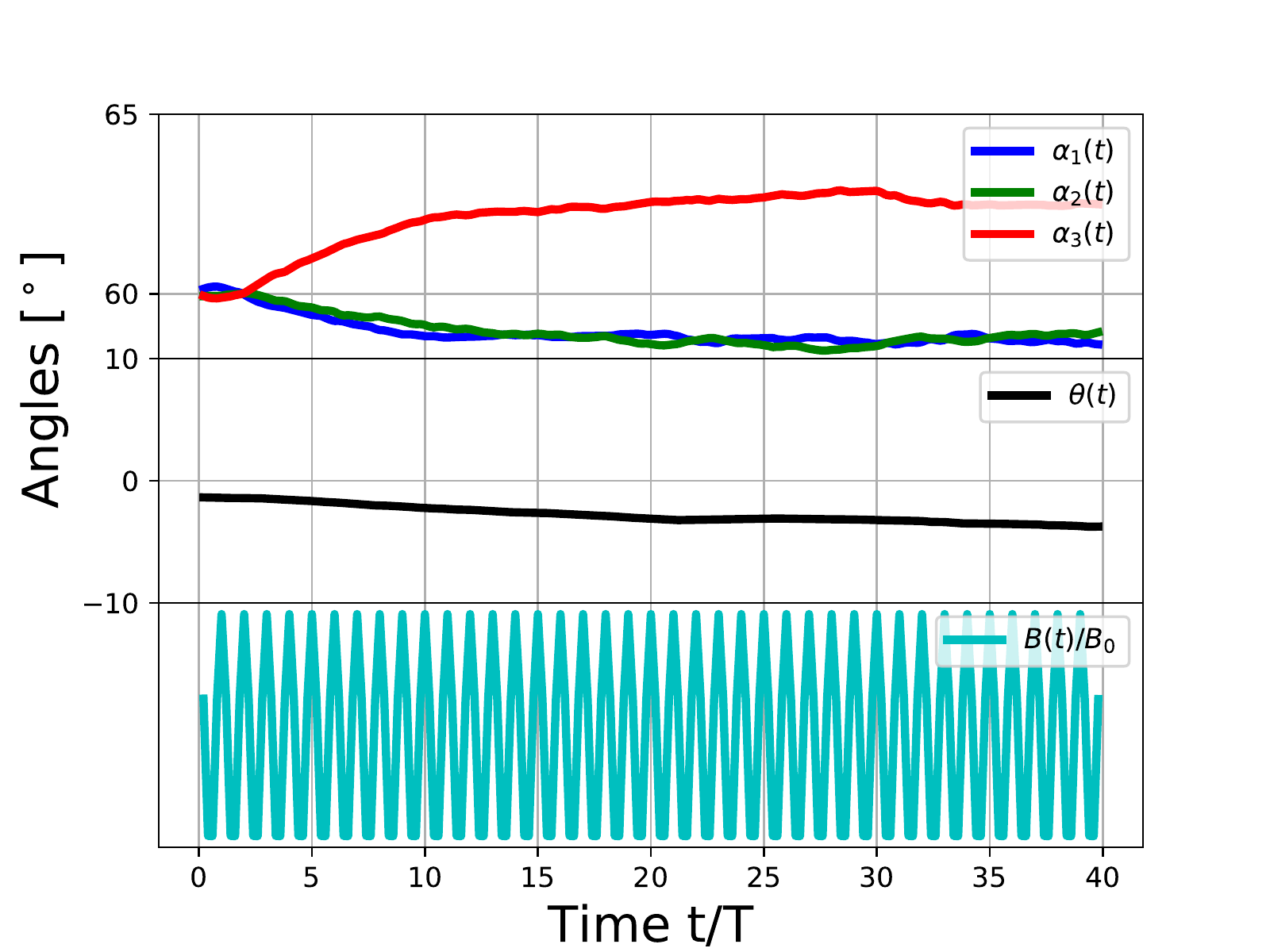}
\caption{High-frequency time propagation of the inner angles $\alpha_i$ within the triangular swimmer and the orientation angle $\theta$ of the swimmer as defined in Fig.~\ref{fig_1}b. LB-parameters: $Bo=0.16$, $L=2.3\times (2R)$, $|B(t)|/|B|=0.36$, $T=125$~$\Delta t$. 
}
\label{fig_2}
\end{figure} 

We introduce the angles $\alpha_i(t)$ between the corresponding arms of the swimmer as shown in Fig.~\ref{fig_1},b as well as the orientation of the swimmer in the plane of the interface $\theta$ defined as the angle between the perpendicular to the line connecting particles 1 and 2 through particle 3 and the z-axis (Fig.~\ref{fig_1}b), where $\theta(t=0)\approx 0$. Starting with an equilateral triangle 
%\MH{Typical distance between the beads in simulations? Should be put somewhere} \AS{It is given in Fig. 2}
($\alpha_i(t_{\mathrm{b}})=60^\circ$, Fig.~\ref{fig_2}, upper panel) it transforms into an isosceles one for $B(t)\neq 0$, while the triangle as a whole only slightly ($< 10^\circ$) rotates after a number of field periods (Fig.~\ref{fig_2}, middle panel). 

\begin{figure*}[htb]
\centering
\includegraphics[width=0.99\textwidth]{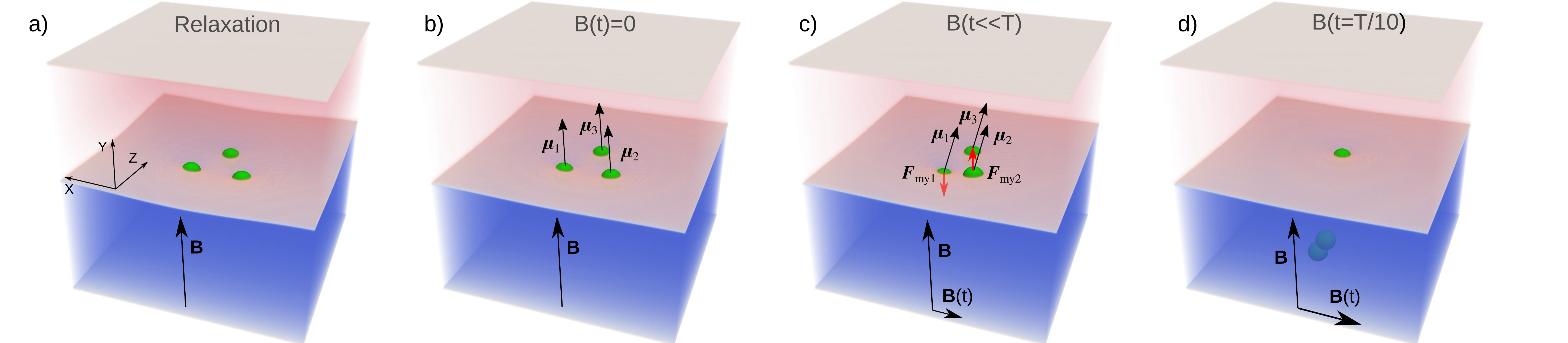}
\caption{Demonstration of the irreversible sinking of particles during the swimmer motion. a) shows the swimmer during its relaxation ($B(t)$ is not applied), b) illustrates the state directly after the driving $B(t)$-field is switched on, c) shows the swimmer just before the sinking of the base particles and d) demonstrates the state of the degraded swimmer, \textit{i.e.} when two particles are detached from the interface. Parameters of simulations: the Bond number $Bo=0.16$, the ratio of magnetic fields $|B(t)|/|B|\approx 0.57$, the period of the external field is $T=10^5$~$\Delta t$.}
\label{fig_3}
\end{figure*}

Approaching to lower frequencies in this regime often leads to a sinking of one or two particles, thus destroying the swimmer. This effect is very pronounced at moderate and large swimmer sizes ($L>3 \times 2R$), hindering the study of its motion at low frequencies. In experiments on magnetocapillary swimmers (refs.~\cite{GrLa15, GrHu16, GrHu17}) sinking of particles was never observed, raising the question of proper parameters in the LB-simulations. Indeed, one can consider the ratio of surface tension and magnetic dipolar forces $F_{\mathrm{st}}/F_{\mathrm{mg}}=2\pi \gamma R/(\mu_0\mu^2/(4\pi r^4_{\mathrm{pp}}))$, where $\mu_0$ is the magnetic permeability of vacuum, $\mu$ is the total bead magnetic moment and $r_{\mathrm{pp}}$ is the distance between the particle centres. Aiming at the maximum of the magnetic force, thus taking $r_{\mathrm{pp}}=2R$, we find the ratio in the experimental situation to be $F_{\mathrm{st}}/F_{\mathrm{mg}}\Big|_{\mathrm{ex}}\approx 10^4$ and in LB-simulations of the order of $F_{\mathrm{st}}/F_{\mathrm{mg}}\Big|_{\mathrm{LB}}\approx 1$. It is obvious from this estimate that in the experiments the surface tension dominates over magnetic interactions, while in the simulations the forces are of the same order. In addition to its strength, the magnetic force is strongly dependent on the mutual orientation of interacting magnetic moments (eq.~(\ref{eq:Fmagn})). In particular, if the particles are in one plane and their magnetic moments are aligned strictly perpendicular to the interface (Fig.~\ref{fig_3}a,b), the out-of-plane magnetic force is zero. If, however, the magnetic moments become tilted by the external time-dependent magnetic field (Fig.~\ref{fig_3}c), the out-of-plane components of the magnetic forces beat the surface tension forces detaching the particles from the interface (Fig.~\ref{fig_3}d). Interestingly, only the particles along the $\vec{B}(t)$-field vector (particles 1 and 2) sink, since by symmetry particle 3 does not experience any out-of-plane magnetic force in this configuration.   

\subsection{Motion at low frequencies}
In order to get closer to the experimental regime and to avoid the sinking of particles, the ratio of the forces $F_{\mathrm{st}}/F_{\mathrm{mg}}$ needs to be increased. A natural way to increase this ratio is by decreasing the magnetic moment. This does not work, however, since it reduces the magnetic repulsion and leads solely to the aggregation of particles.
Alternatively, the surface tension could be increased, 
but the computational effort required to increase the surface tension by several orders of magnitude is prohibitive. 
% \MH{For me, this last sentence raises the question of the validity of the whole LB simulations for the investigations of MC swimmers. This first part of the paragraph should be turned in a more positive way.} 
% \JH{You are absolutely right. Better  now? Alternatives are rare, btw: VOF has problems at the contact line, other diffuse interface methods generally are fine, but one would need a very high reslution of the interface in order to raise the  surface tension as much as needed.
%} 

We therefore numerically set the out-of-plane component of magnetic forces to zero, and thus effectively increase the ratio of surface tension to magnetic force.
This solution is physically sound since all magnetic dipolar forces are pair forces and the total force remains zero. 
%\MH{I'm completely lost here. What is changed? both the magnetic moment and the frequency? If yes, then what happens for high frequency and altered magnetic moment? Also, why ``numerically setting the out-of-plane component of magnetic forces to zero effectively increases the surface tension''? I don't understand the argument. Do you mean that you artificially weaken the magnetic interactions to increase the gap in amplitude between capillary forces and magnetic forces?} 
%\JH{Rephrased. I think it should be clear now.}

\begin{figure}[htb]
\centering
\includegraphics[width=0.48\textwidth]{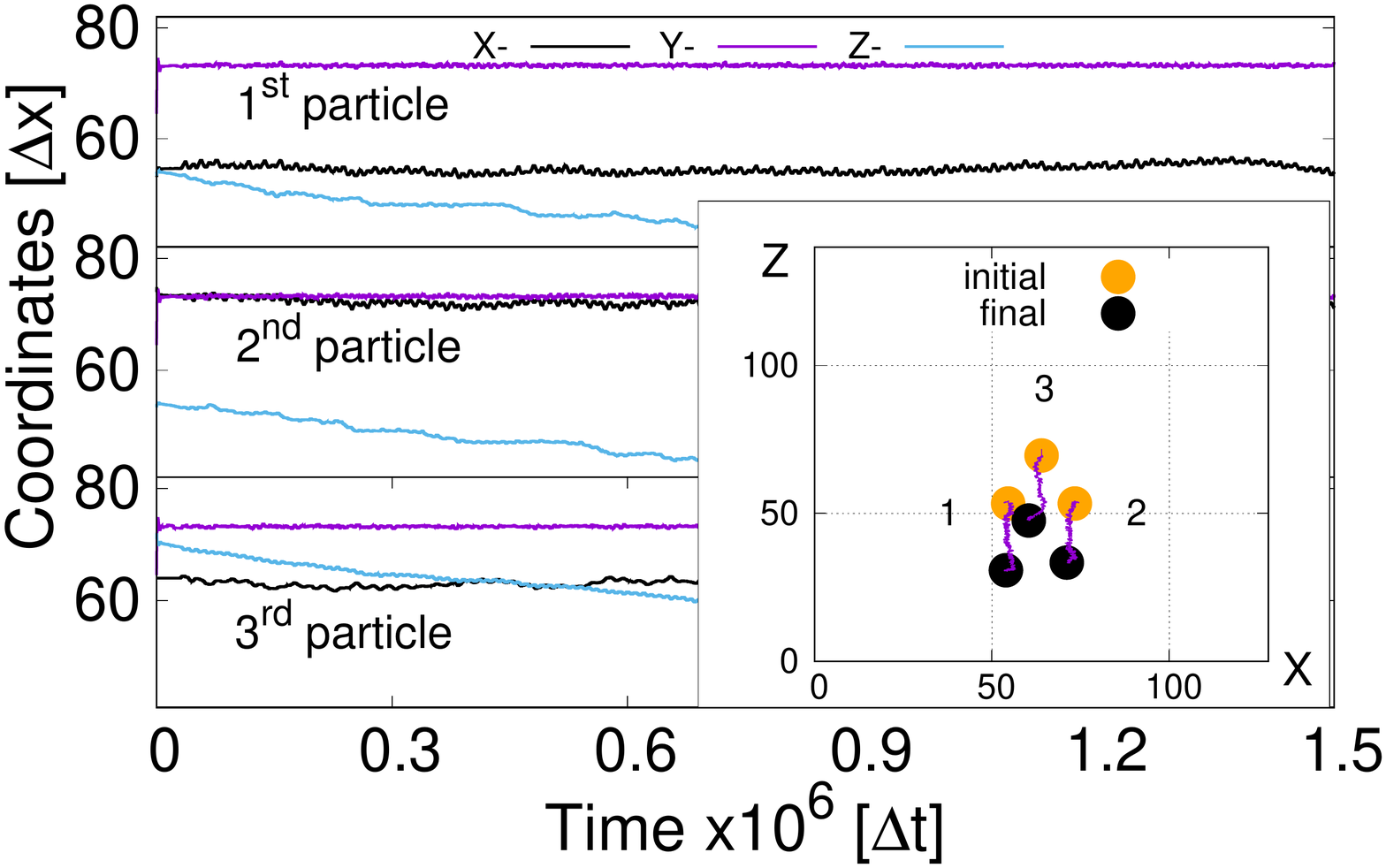}
\caption{Trajectories of each bead during the swimmer motion in the regime of low frequencies. The inset shows
the initial and final positions of the swimmer on the interface. LB-parameters: $Bo=0.16$, $L=1.8\times (2R)$, $|B(t)|/|B|=0.36$, $T=20000$~$\Delta t$.}
\label{fig_4}
\end{figure} 

As shown in Fig.~\ref{fig_4}, the swimmer in this regime propagates in the direction perpendicular to the oscillation of the magnetic field. This aspect is similar to the previously observed motion~\cite{SuZi19}. The way it propagates is, however, different. The amplitudes of particle oscillations are significantly larger than before and reach values around $0.3\times 2R$. Also, all three beads experience pronounced oscillations and not only particles 1 and 2 as it is the case at high frequencies~\cite{SuZi19}.

\begin{figure}[htb]
\centering
\includegraphics[width=0.48\textwidth]{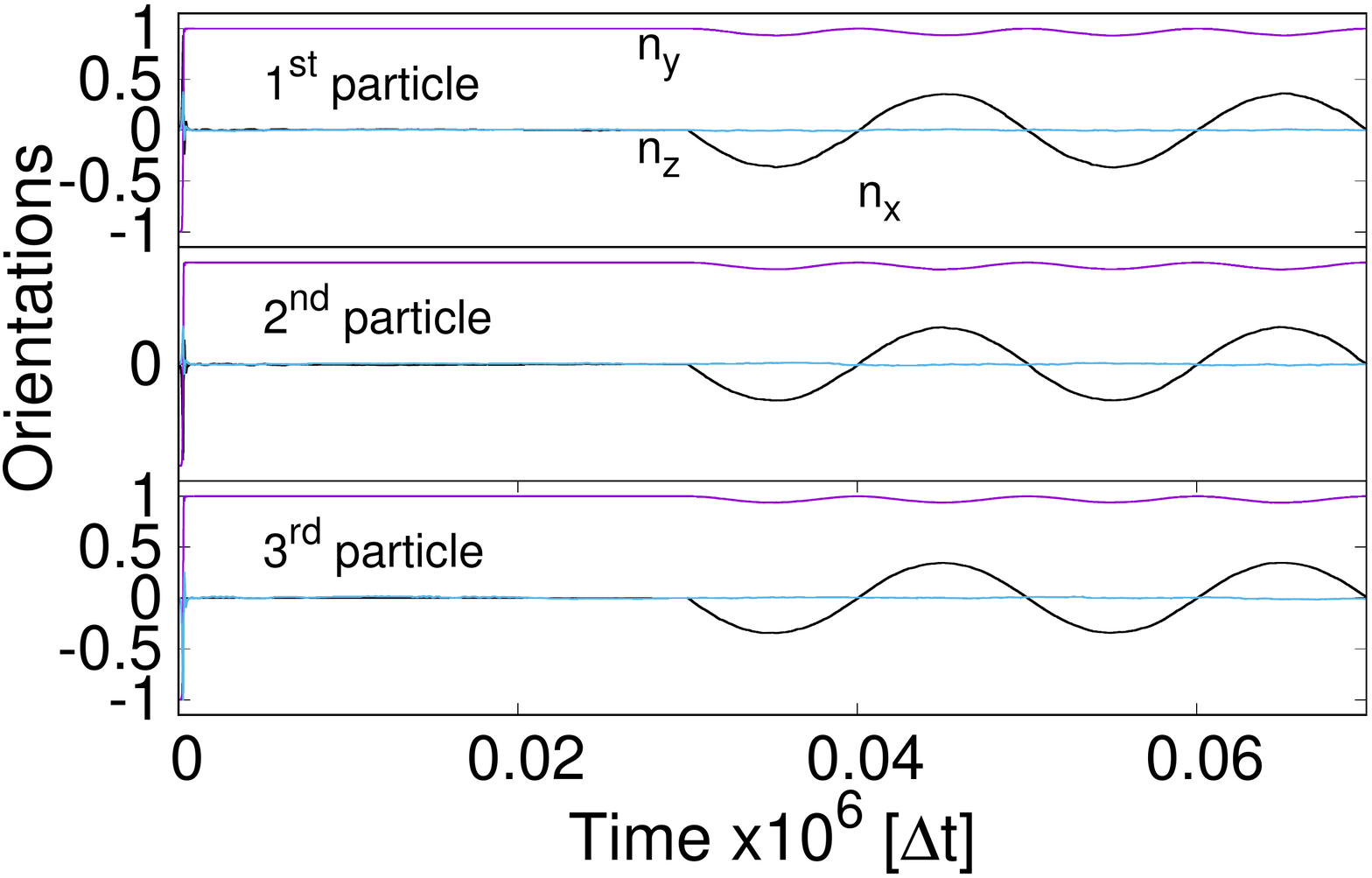}
\caption{Trajectories of orientation vectors $\vec{n}_i$ (Fig.~\ref{fig_1}a) for each bead in the regime of low frequencies. LB-parameters: $Bo=0.16$, $L=1.8\times (2R)$, $|B(t)|/|B|=0.36$, $T=20000$~$\Delta t$. The $B(t)$-field is applied after $t_{\mathrm{b}}=30000$~$\Delta t$.
%\JH{x-axis scaling by 10$^6$ is odd.}\MH{would use 10$^4$ as it compares to the external field period.} \AS{This is done on purpose, since e.g. in Fig. 4 the full simulations range is shown $t=[0;10^6]$.}
}
\label{fig_5}
\end{figure}

Fig.~\ref{fig_5} shows the trajectories of orientation vectors $\vec{n}_i$ stressing the fact that orientations of the particles follow the direction of the time-dependent external magnetic field which is applied after a relaxation time of $t_{\mathrm{b}}=30000$~$\Delta t$. As mentioned in ref.~\cite{SuZi19}, the time $t_{\mathrm{b}}$ is chosen after studying vertical relaxations of single and multiple particles at the fluid-fluid interface. It assures that for $t>t_{\mathrm{b}}$ the vertical motion of both the particles and the fluid is negligibly small. The maximum declination of the direction vector $n_{\mathrm{x} i}^{\mathrm{max}}\approx 0.36$ is the consequence of the applied time-dependent and static magnetic fields $|B(t)|/|B|\approx 0.36$. Since direction vectors are unit vectors $|\vec{n}_i|=1$ and the magnetic fields are applied in the xy-plane, the $n_{\mathrm{z} i}$-component remains nearly zero attaining the maximum declination for $\Delta n_{\mathrm{y} i}=1-\sqrt{1-n_{\mathrm{x} i}^2}\approx 0.07$ (see Fig.~\ref{fig_5} for $n_{\mathrm{y} i}$).   

\begin{figure}[htb]
\centering
\includegraphics[width=0.48\textwidth]{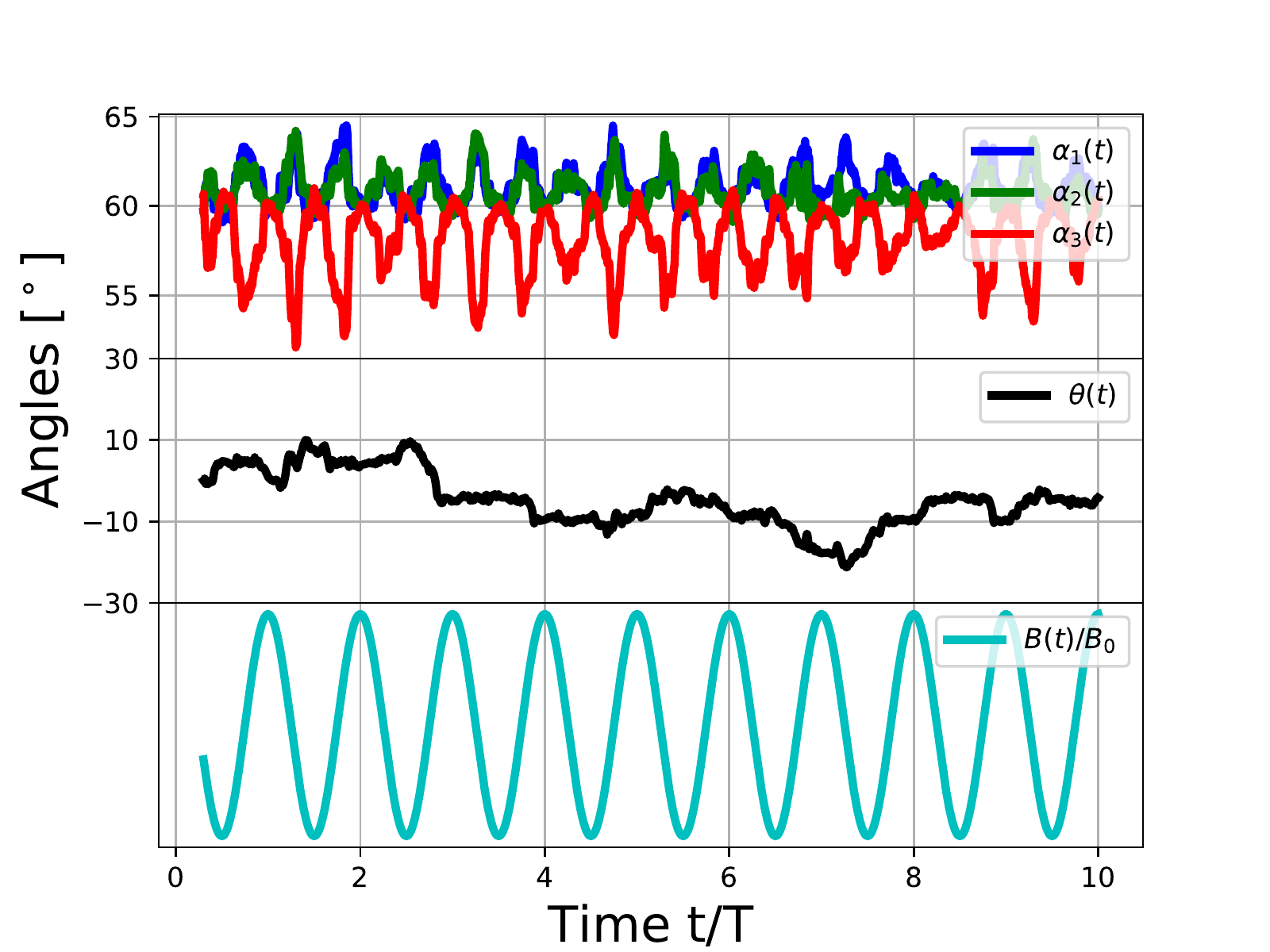}
\caption{Time propagation of the inner angles $\alpha_i$ within the triangular swimmer and the orientation angle $\theta$ of the swimmer as defined in Fig.~\ref{fig_1}b. LB-parameters: $Bo=0.16$, $L=1.5\times (2R)$, $|B(t)|/|B|=0.36$, $T=100000$~$\Delta t$.}
\label{fig_6}
\end{figure}

The orientation of the swimmer $\theta$ in this case does not show any regular pattern (Fig.~\ref{fig_6}), but it indicates an overall slight rotation of the triangle with respect to the initial orientation by less than $10^\circ$. Since during the motion of the swimmer $\theta$ reaches values exceeding $20^\circ$, we conclude that the swimmer in this regime tries to synchronize its orientation with respect to the driving $B(t)$-field. 
%\GG{Any idea why the rotation fails to synchronize with the field? Is it transient or does it never synchronize?} \AS{As long as the magnetic moment is paramagnetic, $\theta$ does not synchronize in the simulations.} %
Fig.~\ref{fig_6} demonstrates significant differences in the time dependence of angles $\alpha_i$ compared to the high frequency mode. Here, angle $\alpha_3$ associated with the third particle periodically decreases while angles $\alpha_1$ and $\alpha_2$ increase to the same amount. This is the consequence of the reduced magnetic moments $\vec{\mu}_1$ and $\vec{\mu}_2$ on the average of the field period, hence a reduced magnetic repulsion between the particles. As a result, the attractive capillary interaction pushes particles 1 and 2 closer to each other compared to the situation in equilibrium ($\alpha_i=60^\circ$). 

The fact that the frequencies of angle oscillations are two times higher than the ones related to the driving magnetic field reflects the magnetic pair-interaction nature. Indeed, the induced magnetic moment can be written as $\vec{\mu}=\chi V_0/\mu_0 (\vec{B}+\vec{B}(t))$ or simply $\mu\sim \mathrm{const}+\cos \omega t$, where $\chi$ is the magnetic susceptibility and $V_0$ is the volume of a particle. If the magnetic moments are oriented nearly perpendicular to the interface, then the magnetic repulsion force scales as $F_{12}\sim (\vec{\mu}\cdot \vec{\mu})=(\mathrm{const} + C_1 \cos\omega t + C_2 \cos 2\omega t)$, where $C_{1, 2}$ are some physical constants. In other words, the second harmonics are inherent in magnetic interactions and since each particle interacts with two others at the same time, several first and second harmonics are always present in their trajectories with different weights. 

\begin{figure}[htb]
\centering
\includegraphics[width=0.48\textwidth]{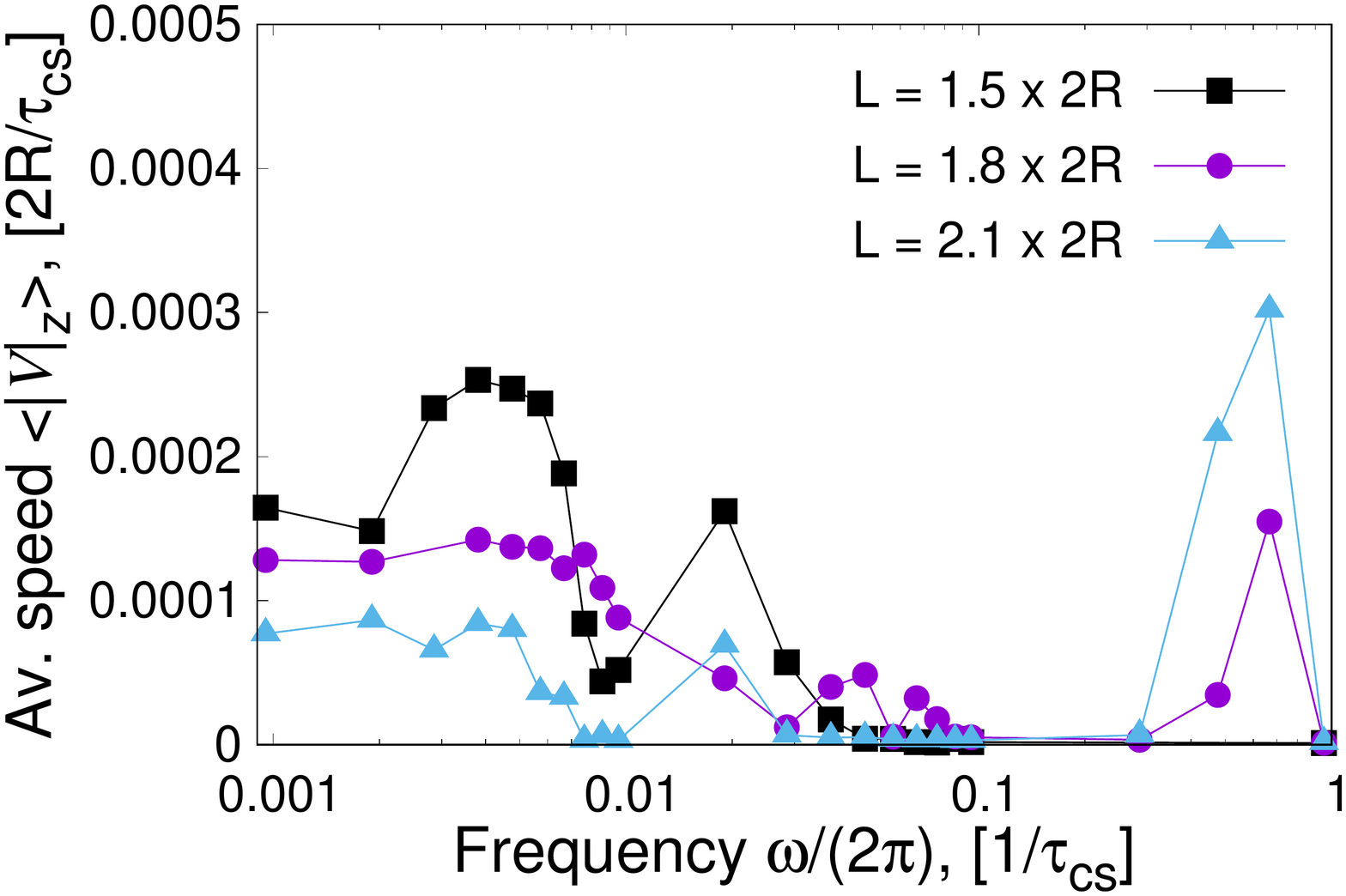}
\caption{Speed of the centre of mass of the swimmer averaged over multiple periods vs. frequency of the external magnetic field in the regime of low frequencies. The swimmer propagates mainly along the z-axis, a drift along the x-axis can be ignored. LB-parameters: $Bo=0.16$, $|B(t)|/|B|=0.36$.}
\label{fig_7}
\end{figure} 

Fig.~\ref{fig_7} summarizes the behaviour of the average velocity of the swimmer as a function of the frequency of the external field $B(t)$ for different swimmer sizes $L$. In general, swimmer velocities are sensitive variables. Therefore, the swimming velocity results from averaging in time and on different numerical trajectories, namely: first, the swimmer velocity is calculated according to the definition given by eq.~(\ref{eq_v_swim}) which describes a time average of the displacement of the swimmer's centre of mass. Second, for every swimmer size $L$ we average over two trajectories differing in the time when the $B(t)$ is applied, \textit{i.e.} $t_{\mathrm{b}}=30000$~$\Delta t$ and $t_{\mathrm{b}}=100000$~$\Delta t$, meaning that initial positions of the swimmer in both cases are slightly different. 
This averaging is required to smoothen the discretisation effects of the particles and the interface. 
Compared to the similar dependence of the regime at high frequencies (Fig.~9a in ref.~\cite{SuZi19}), we clearly identify the swimmer operation in a much broader range of frequencies for all swimmer sizes.

%\MH{Next paragraph: I'd be very careful in the following regarding the role of inertia in the dynamics. Most readers reject inertia as soon as the words ``low RE'' are used in a paper. From my point of view, the whole paragraph treats the beads inertia as a well and widely know ingredient of the dynamics. }

Fig.~\ref{fig_7} also provides a deep insight into the physics of the swimmer motion at low and moderate frequencies of its driving. It captures the whole complexity of the motion in terms of capillary and magnetic interactions (magnetocapillary potential), hydrodynamic interactions, the behaviour of the interface, triangular geometry of the swimmer and the effects associated with the inertia of the particles. Although there is a number of studies dealing with the physics of swimmer motion~\cite{NaGo04, GoAj08, Feld06, ChLa15, RiFa18, ZiHu19, HuTr20}, it is hardly possible to include all the aforementioned effects in a single theoretical formalism. Studies relying on the force-based approach 
%do not take the effect of particle inertia into account. They 
suggest that the maximum swimmer velocity should be centered around the frequencies associated with the harmonic potential controlling the arm length, \textit{e.g.} $\omega_{\mathrm{St}}=k/(6\pi \eta R)$~\cite{RiFa18, ZiHu19}.
At the same time, a naive estimate related to the resonance frequency a harmonic oscillator is known to scale $\omega_{\mathrm{spring}} \sim \sqrt{k/m}$. Taking the approximate values of the spring constants $k_{12 x}$ extracted from the particle trajectories of the swimmer of moderate sizes (Appendix A), we estimate for the maximum velocities the range $\omega (\langle V\rangle^{\mathrm{max}}) \approx [0.0005 ; 0.009]$~$1/\tau_{\mathrm{cs}}$. Fig.~\ref{fig_7} confirms this behaviour by showing broad velocity distributions at different swimmer sizes in this range of frequencies. A decay of averaged velocities as a function of the swimmer size $L$ in this $\omega$-range reflects decreasing hydrodynamic interactions upon the swimmer growing, which is shown by expression (43) of ref.~\cite{ZiHu19} though in the Oseen-tensor representation only ($R/L<1/6$). Its Rotne-Prager extension applicable to the sizes considered here ($1/4<R/L<1/3$) confirms the observed behaviour of $\langle V^{\mathrm{max}}(L)\rangle$ for $\omega \in [0.001; 0.01]$~$1/\tau_{\mathrm{cs}}$.

\subsection{Motion at low frequencies and finite internal magnetic moment}
The simulated triangular magnetocapillary swimmer presented so far shows how its motion differs depending on the applied field frequency $\omega$, properties of the interface or the swimmer size $L$. Compared to the experimental situation in which high in-plane cyclic bead rotations are observed (Fig.~6 in ref.~\cite{GrLa15}), the simulated swimmer never shows such type of motion since the propagation of $\theta$ in Fig.~\ref{fig_6} is not periodic. 
Indeed, the strong in-plane bead rotations observed in the experiments point to more complex magnetic properties of the beads.

For unraveling the magnetic properties of the beads a series of experiments was performed using a single particle placed at the interface and driven by an external magnetic field~\cite{GrHu19}. Therein, the magnetic bead rotates under the application of a constant magnetic field in the plane of the interface when the field rapidly changes its orientation by $180^\circ$. Assuming that the magnetic moment is of paramagnetic nature, \textit{i.e.} $\vec{\mu} \sim \vec{B}$, the associated magnetic torque on the particle should be zero ($\vec{T}\sim [\vec{\mu} \times \vec{B}]=0$) and cannot cause the particle to rotate around its own axes. This fact leads to the hypothesis of the existence of a permanent internal magnetic moment that is randomly oriented when the particle is placed at the interface. Upon switching the field orientation from "$+$" to "$-$" the permanent internal magnetic moment follows the field and mechanically rotates the particle. Moreover, a correct linear scaling was experimentally observed for the maximum rotation frequency of the bead with respect to the magnetic field amplitude (Fig.~8 in ref.~\cite{GrHu19}). 

The origin of the small constant internal magnetic contribution in the particles is still under debate~\cite{GrHu19}. Taking into account their size (diameters of several hundred micrometers) and almost perfect spherical form, it can be shown by exact numerical micromagnetic simulations (sect.~5 in ref.~\cite{GrHu19}) that their net magnetic moment should be zero in the absence of an external field. The latter should also be true in much larger systems, \textit{i.e.} above diameters 1-3 $\mu$m for which the micromagnetic simulations were performed. At large particle sizes ($> 1\mu$m) long-ranged magnetic dipolar interactions start favoring the formation of magnetic domains that are randomly oriented in space and their number grows upon reaching hundreds of micrometers. Considering that the particles are highly monodisperse in density \cite{GrHu19}, only two effects can cause the presence of a finite internal magnetic moment: \textit{i)} defects at the boundaries of some magnetic domains (similar to the \textit{Barkhausen effect}) and/or \textit{ii)} the fabrication process of the magnetic beads. In the latter case, steel wires are originally cut into small cylinders, then pressed into spherical dies and finally rounded~\cite{GrHu19}. We speculate that this process might induce additional magnetic anisotropies in the particles.

\begin{figure}[htb]
\centering
\includegraphics[width=0.48\textwidth]{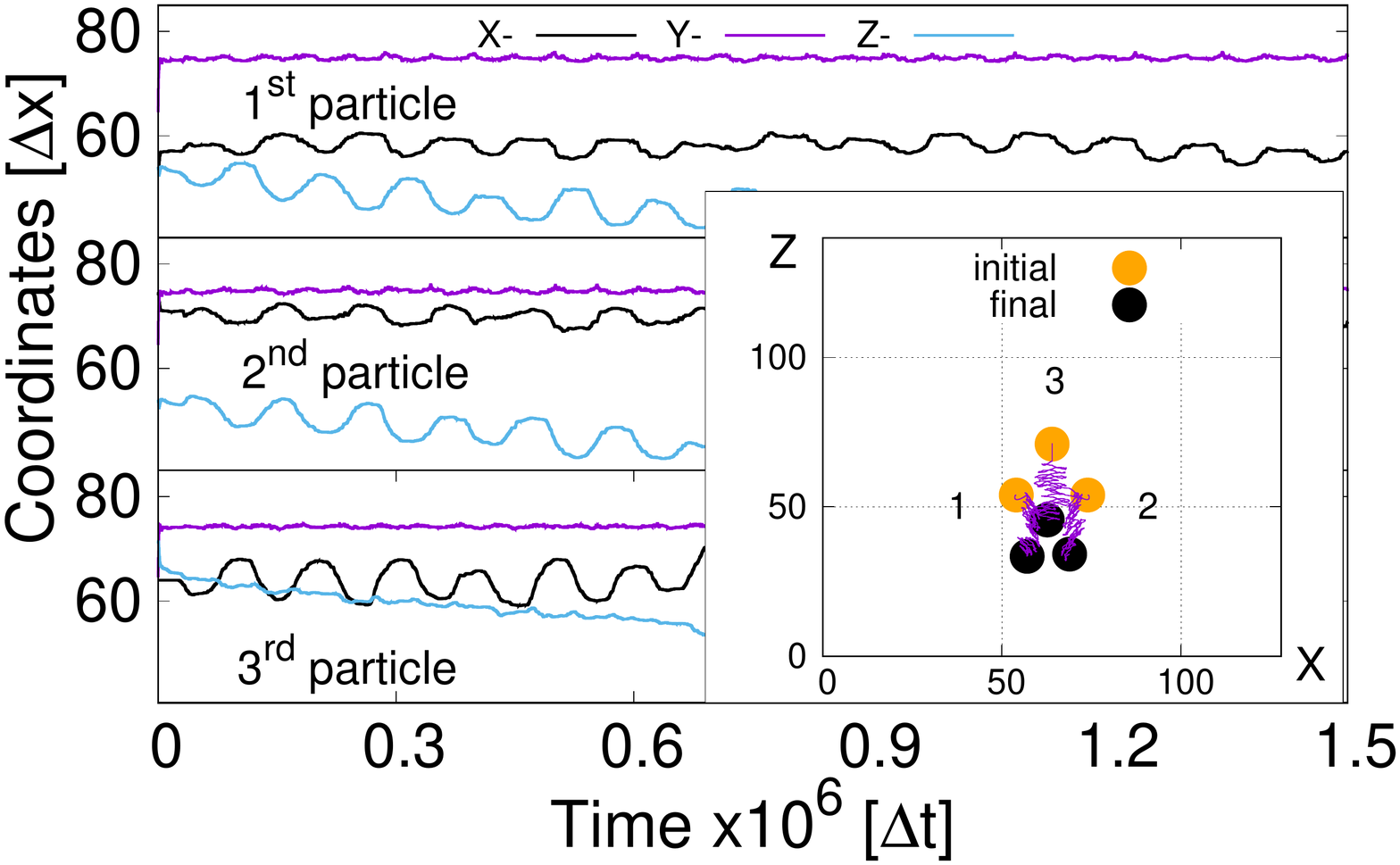}
\caption{Trajectories of each bead during the swimmer motion in the regime of low frequencies and finite internal magnetic moment. The inset shows the initial and final positions of the swimmer on the interface. LB-parameters: $Bo=0.16$, $L=1.5\times (2R)$, $|B(t)|/|B|=0.36$, $T=100000$~$\Delta t$, $\mu\ind{int x}=0.1\mu\ind{ind}$, $\mu\ind{int y}=0$, $\mu\ind{int z}=0$.} 
\label{fig_8}
\end{figure} 

\begin{figure}[htb]
\centering
\includegraphics[width=0.48\textwidth]{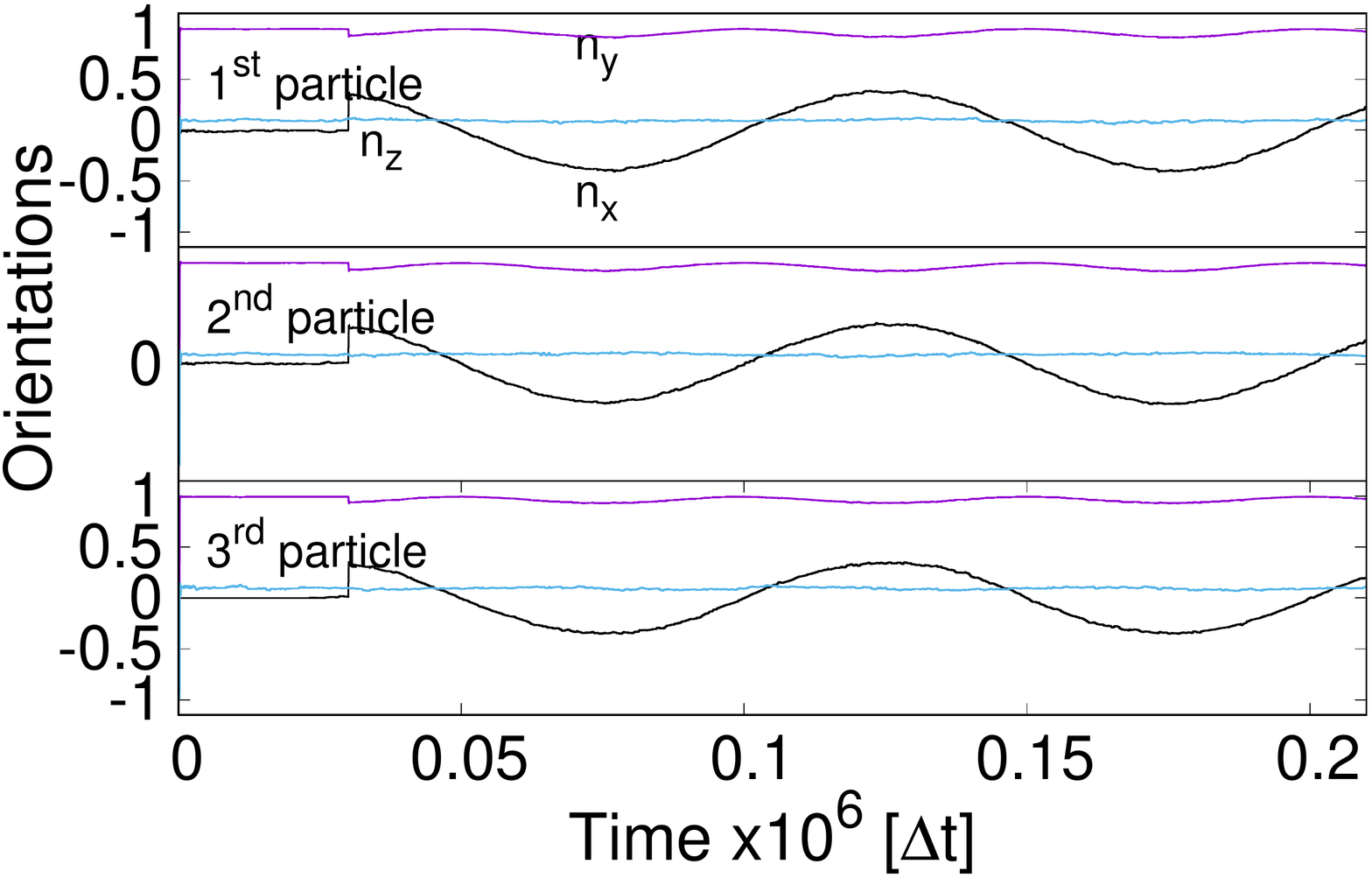}
\caption{Trajectories of orientation vectors $\vec{n}_i$ (Fig.~\ref{fig_1}a) for each bead in the regime of low frequencies and finite internal magnetic moment. LB-parameters: $Bo=0.16$, $L=1.5\times (2R)$, $|B(t)|/|B|=0.36$, $T=100000$~$\Delta t$, $\mu\ind{int x}=0.1\mu\ind{ind}$, $\mu\ind{int y}=0$, $\mu\ind{int z}=0$. The $B(t)$-field is applied after $t_{\mathrm{b}}=30000$~$\Delta t$. 
%\MH{How are the beads rotating? I still don't understand. Also, does the vector $\vec{n}_i$ includes the two magnetic contributions to the magnetic moment?  Careful here, it seems that the beads are only rotating out of plane when everywhere in the paper we read that the beads rotate in plane in experiments}\AS{Not the beads rotate, but the swimmer reorients itself in the plane.}
}
\label{fig_9}
\end{figure} 

Being equipped with the experimental proof for the existence of the permanent internal magnetic moment, we assume that the total magnetic moment in each particle has the \textit{induced} $\vec{\mu}_{\mathrm{ind} i}(\vec{B})$ and the \textit{internal} constant $\vec{\mu}_{\mathrm{int} i}$ magnetic contributions
\begin{equation}
\displaystyle \vec{\mu}\ind{tot}_i(\vec{B}) = \vec{\mu}\ind{ind}_i(\vec{B}) + \vec{\mu}\ind{int}_i,
\label{eq_mu_tot}
\end{equation} 
where $\vec{\mu}\ind{ind}_i(\vec{B})=\chi V_0 \vec{B}/\mu_0$ is the function of the external field $\vec{B}$ in direction and amplitude, while $\vec{\mu}\ind{int}_i$ is fixed in its direction $\vec{n}_i$ and strength irrespective of the $\vec{B}$-orientation.

Appendix B provides full details of how magnetic forces and torques are modified if eq. (\ref{eq_mu_tot}) holds. In particular, magnetic forces gain three additional terms, since internal magnetic moments interact with the induced ones and with themselves in different particles. The same applies to the magnetic torques. For the strength of the internal magnetic moment we rely on experimental observations~\cite{GrHu19}, where the strength was estimated to be approximately in the range $|\mu\ind{int}|\in [0.1; 0.15]|\mu^{\mathrm{max}}_{\mathrm{ind}}|$ 
%\AS{
of the maximum induced magnetic moment. 
% }\MH{According to this definition, the internal moment should be time dependant as the induced one is.} 

Fig.~\ref{fig_8} represents the motion of the swimmer at low frequencies and in the presence of finite internal magnetic moment $\mu\ind{int x}=0.1 \mu\ind{ind}$ ($\mu\ind{int y}=0$, $\mu\ind{int z}=0$). Each of the three particles have the same internal magnetic moment in strength and direction and the induced magnetic moment is present as described above. There is a notable difference in the way how all particles move in this case with respect to previous modes (Fig.~\ref{fig_4}): both x- and z-components of the particles along which the $B(t)$-magnetic field is applied experience sizable oscillations while the top particle performs oscillations only along the x-direction. The net swimmer displacement in this regime is approximately the same as in the absence of the internal magnetic moment (inset of Fig.~\ref{fig_8}).   

The dynamics of rotation vectors $\vec{n}_i$ ($\mu\ind{int x}=0.1 \mu\ind{ind}$, Fig.~\ref{fig_9}) does not essentially differ from that when the internal magnetic moment is zero (Fig.~\ref{fig_5}). We only witness a very tiny z-component of $\vec{n}_i$ for all particles which is attributed to a new in-plane magnetic equilibrium due to the presence of $\mu\ind{int x}$.

\begin{figure}[htb]
\centering
\includegraphics[width=0.48\textwidth]{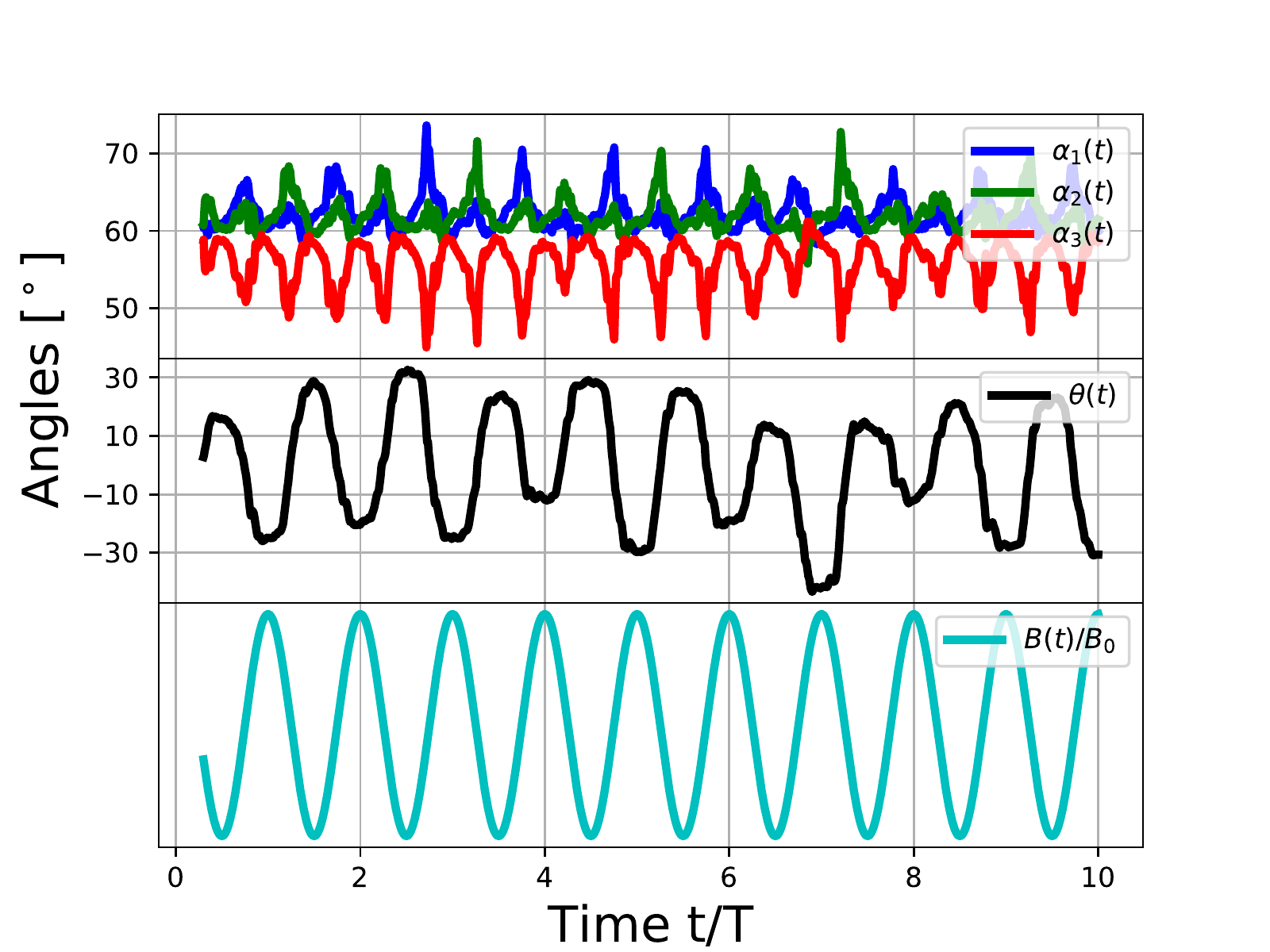}
\caption{Time propagation of the inner angles $\alpha_i$ within the triangular swimmer and the orientation angle $\theta$ of the swimmer as defined in Fig.~\ref{fig_1}b. LB-parameters: $Bo=0.16$, $L=1.5\times (2R)$, $|B(t)|/|B|=0.36$, $T=100000$~$\Delta t$, $\mu\ind{int x}=0.1\mu\ind{ind}$, $\mu\ind{int y}=0$, $\mu\ind{int z}=0$.}
\label{fig_10}
\end{figure}

The presence of the finite internal magnetic moment leads to substantial in-plane dynamics of all the beads within the swimmer. As shown in Fig.~\ref{fig_10}, the orientation of the swimmer $\theta$ follows exactly the period of the external magnetic field $B(t)$ and on the large time scale the swimmer keeps its in-plane orientation such that $\int_0^{t\rightarrow \infty}\theta (t)dt\approx 0$. Additionally, we observe that the strength of $\mu\ind{int}$ defines how strong the angle $\theta$ deviates from the equilibrium $\theta=0$ meaning that one can judge about the magnitude of $\mu\ind{int}$ based on $\langle\theta^{\mathrm{max}}\rangle$. Along with the pronounced $\theta(t)$-dependence we detect several changes in the propagation of $\alpha_i(t)$. In contrast to the triangle deformations in the absence of $\mu\ind{int}$ (Fig.~\ref{fig_6}), where all angles $\Delta \alpha_i^{\mathrm{max}}<5^\circ$, we now notice larger triangle deformations $\Delta \alpha_i^{\mathrm{max}}\approx 10^\circ$ that again depend on the magnitude of $\mu\ind{int}$. Moreover, an asymmetric propagation of the base angles $\alpha_1$ and $\alpha_2$ (Fig.~\ref{fig_10}, upper panel) complies with the asymmetry introduced by the direction of the internal magnetic moment $\mu\ind{int x}$: the x-components of the induced and the internal magnetic moments are aligned anti- or -parallel depending on the $B(t)$-direction.

Noteworthy is also the orientation of the internal magnetic moment. The dynamics presented in Fig.~\ref{fig_10} is valid for the x- or in general an in-plane component of $\mu\ind{int}$. Once one introduces $\mu\ind{int y}$ or an out-of-plane component solely, which is additive to the induced magnetic moment, we do not observe periodic reorientations of $\theta$ (not shown here).

Although quantitatively there might be differences between our simulations and the experimental observations for angles $\alpha_i$ and $\theta$ (\textit{e.g.} Fig.~6 in ref. \cite{GrLa15}), qualitatively we recover the main experimental findings: when introducing a smaller in-plane component of the internal magnetic moment $\mu\ind{int}\approx 0.1\mu\ind{ind}$, the swimmer demonstrates remarkable reorientations defined by the angle $\theta$ which follow the external $B(t)$-field by its simultaneous swift propulsion at the interface.   

\begin{figure}[htb]
\centering
\includegraphics[width=0.48\textwidth]{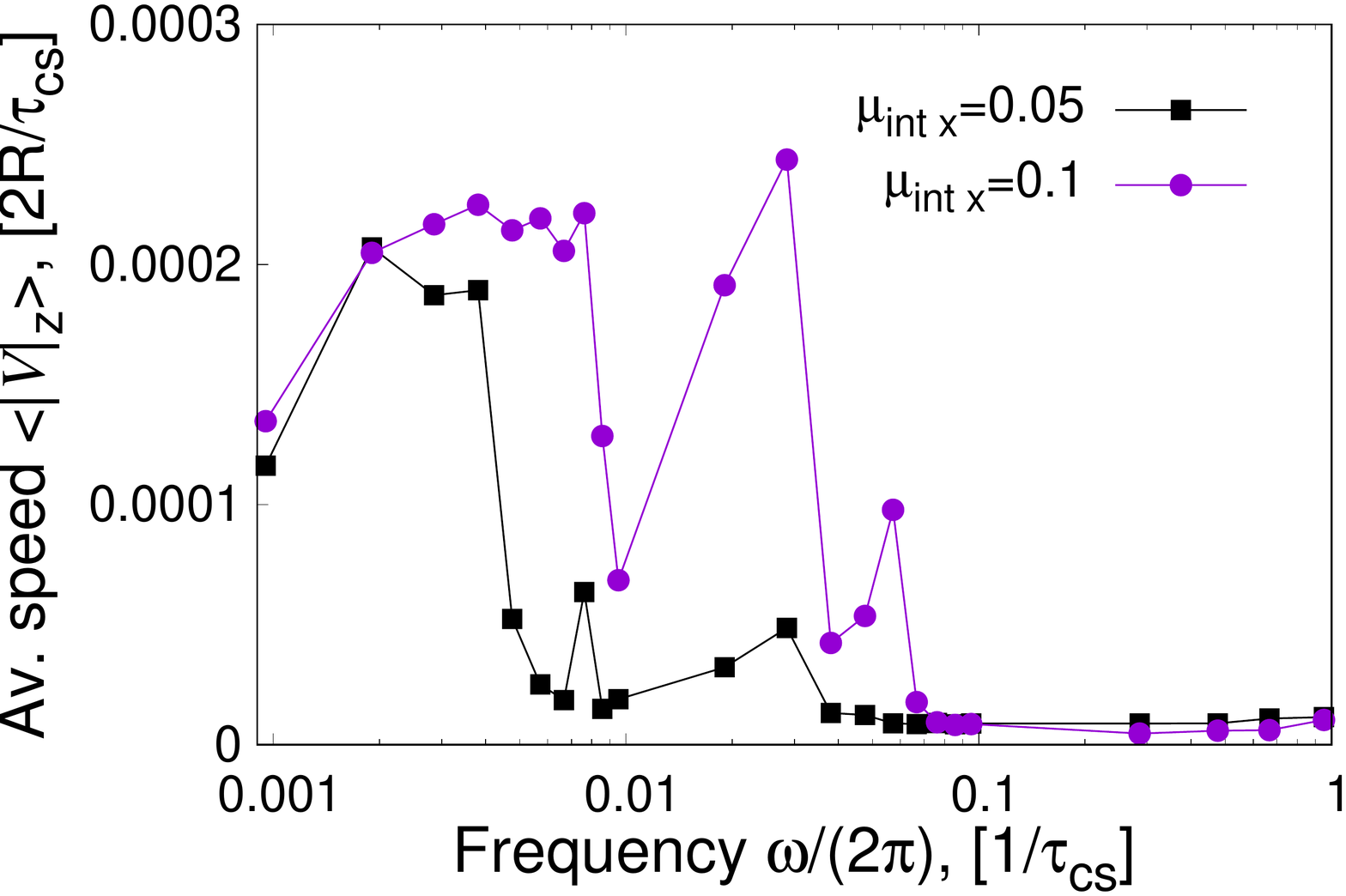}
\caption{Speed of the centre of mass of the swimmer averaged over multiple periods vs. frequency of the external magnetic field in the regime of low frequencies and finite internal magnetic moment. LB-parameters: $Bo=0.16$, $L=1.5\times (2R)$, $|B(t)|/|B|=0.36$.}
\label{fig_11}
\end{figure}

Finally, the dependence of the averaged velocity on the applied frequency (Fig.~\ref{fig_11}) does not change significantly with respect to the situation with the absent internal magnetic moment (Fig.~\ref{fig_7}), \textit{i.e.} the swimmer is most efficient for
frequencies $\omega/(2\pi)\sim [0.001; 0.01]$~$1/\tau_\ind{cs}$. 
The increasing value of the magnetic moment $\mu\ind{int x}$ broadens the averaged velocity as expected.

\section{Summary and Discussion}
%We note that the word \textit{regime} in this study should be understood %in the context of both physics and the employed numerical method. It means that both physical parameters and abilities of the method define the type of the observed swimmer motion. 
%\JH{Isn't this a trivial statement?}

\subsection{Different regimes of motion}

Using the lattice Boltzmann method with the Shan-Chen model for the fluid-fluid-interface we demonstrate three different regimes of stable swimmer motion: the regime with paramagnetic particles at high \textit{i)} and low \textit{ii)} frequencies and \textit{iii)} the regime of ferromagnetic particles at low frequencies.

In regime \textit{i)} (ref.~\cite{SuZi19}) the magnetic moments of all particles are induced by a set of externally applied static and oscillating magnetic fields. The swimmer propagates having small particle displacements and shows neither typical sizable in-plane rotations of the beads as observed in the experiments~\cite{GrLa15} nor periodic reorientations of the swimmer (evolution of $\theta$ in Fig.~\ref{fig_2}). The peaks of the averaged swimmer velocity (Fig.~11 in ref.~\cite{SuZi19}) are observed at high frequencies (in lattice units) characteristic to viscous or coasting times of the particles. Reduction of the driving frequency required for a better temporal resolution of the motion often led to sinking of one or several particles (Fig.~\ref{fig_2}), thus destroying the swimmer. We find that sinking is caused by sizable out-of-plane components of magnetic forces exceeding the surface tension force.

Regime \textit{ii)} is achieved through suppressing the vertical component of the magnetic force of the swimmer having otherwise the parameters of regime \textit{i)}. As a result, the swimmer is capable to propagate at significantly lower frequencies associated with the magnetocapillary potential strength and a good temporal resolution of its motion is accomplished. The swimmer in this regime shows sizable side deformations, however, no periodic reorientations characterised by the angle $\theta$ (Fig.~\ref{fig_6}). 

Finally, regime \textit{iii)} is primarily characterised through the existence of an additional small constant internal magnetic contribution that is evidenced in the experiments~\cite{GrHu19}. The swimmer demonstrates a motion at low characteristic frequencies and possesses typical $\theta$-reorientations (Fig.~\ref{fig_10}) similar to those observed experimentally \cite{GrLa15}. It should be noted that only the in-plane component of the internal magnetic moment causes the typical swimmer motion seen experimentally, while the out-of-plane magnetic contributions do not lead to any sizable swimmer reorientations, since in this case it adds to the vertical paramagnetic moment.  

In regimes \textit{ii)} and \textit{iii)} we witness one remarkable non-trivial effect associated with the frequency of the driving field. Even if the external magnetic field $B(t)$ has only one frequency $\omega$, the associated magnetic force might have a second harmonics, since the magnetic force between each pair of particles scales $F_{ij}\sim (\vec{\mu}_i\cdot \vec{\mu}_j)\sim \mathrm{const}+\cos\omega t +\cos 2\omega t$  for each magnetic moment $\vec{\mu}_i \sim (\vec{B}+\vec{B}(t))$. This frequency doubling effect is clearly seen in Figs.~\ref{fig_6} and \ref{fig_10}. This should be kept in mind in case of magnetocapillary or in general any magnetically driven swimmer, when applying theoretical \textit{e.g.} bead-spring models: the force always follows $ 2\omega $ although the field is applied with frequency $\omega$.

\subsection{Swimmer velocities}

%We compare the velocities of experimental and simulated magnetocapillary %swimmers. 
In experiments, for bead diameters $2R=500$~$\mu$m the average swimmer velocity reaches values up to $\langle V\ind{exp}\rangle\approx 0.3$~$(2R)/T$~\cite{GrLa15} for the ratio of oscillating to static field $|B(t)|/B\approx 0.5$ and about $\langle V\ind{exp}\rangle\approx 0.02$~$(2R)/T$ for moderate $|B(t)|/B\approx 0.1$. Our LB-simulations yield for the maximum average velocities $\langle V\ind{LB}\rangle$ 
$\approx 0.0004$~$(2R)/T$ in regime \textit{i)}~\cite{SuZi19} and approximately $\langle V\ind{LB}\rangle\approx 0.06$~$(2R)/T$ in both regimes \textit{ii)} and \textit{iii)} (sects. 3.2 and 3.3, respectively). Although the simulated velocities in absolute units are of the same order of magnitude in all the described regimes ($\langle V\ind{LB}\rangle\approx 10^{-5}$ l.u.), we reach a better agreement with the experiment in units of $(2R)/T$ for regimes \textit{ii)} and \textit{iii)}. It is also in line with the analytical predictions for the triangular swimmer velocity in ref.~\cite{ZiHu19} (eq.~(43)) or for a dumbbell swimmer including effects of inertia (ref.~\cite{HuTr20}, eq.~(2)): the lower the potential constant is (estimates in Appendix A yield $k\approx 10^{-5}$~l.u.), the lower are the frequencies of the peak velocities leading to a better time resolution and the higher are the maximum velocity amplitudes. Finally, we note that since the swimmer velocity typically scales quadratically $\langle V\rangle \sim A^2$~\cite{PaMe17, RiFa18, ZiHu19, HuTr20} with the external driving amplitude $A$, this is also the way to tune up the velocity. With triangular magnetocapillary swimmers it has, however, a limitation at increasing field ratios, since at values $|B(t)|/B\approx 0.6$ a dynamic transition from a triangular to a linear swimmer configuration occurs (Fig.~2 of ref.~\cite{GrHu17}). We are able to reproduce this transition in our LB-simulations and therefore fix the ratio around $|B(t)|/B\approx 0.36$ to assure the triangular form~\cite{SuZi19}.  

\subsection{Simulation method and parameters}

The presented simulations of magnetocapillary swimmers are a challenging task. On the one hand, we model the fluid-fluid interface and its dynamics coupled with the dynamics of the externally driven magnetic particles. On the other hand, the magnetic properties of the swimmer are included in the simulation by taking into account not simply effective external repulsive forces but rather detailed paramagnetic and ferromagnetic contributions to the total magnetic moment of the beads leading to the particle repulsion. As a consequence, such thorough modeling of the problem allows for very detailed insights into the static properties of magnetocapillary swimmers such as horizontal and vertical positioning of the beads upon swimmer self-assembling, the conditions for which the particles may detach from the interface and a realistic description of capillary phenomena for finite particle sizes and moderate inter-particle distances~\cite{SuZi19}. Moreover, the rich physics of the swimmer propagation associated with the potential strength and the interface dynamics is reflected in their velocity-vs-frequency dependencies (Figs.~\ref{fig_7} and \ref{fig_11}). 

At the same time, the choice of the method and the limitation in reaching realistic surface tensions with acceptable computational effort is also responsible for the sinking of particles upon the swimmer motion (Fig.~\ref{fig_3}). It helps better understand the experimental conditions such as a very high surface tension that practically pins the floating particles to the interface permitting thus only in-plane particle dynamics. A possible solution of the problem associated with the sinking of beads in LB-simulations consists in modifying not the interface, but rather in setting the vertical components of magnetic forces to zero, thus suppressing the out-of-plane swimmer dynamics. 

%For the present parameters ($R$, the size of the simulation box) a direct increase of the surface tension is limited by the maximum force one can apply to a lattice node by keeping the simulation stable. The limitation can be circumvented by increasing the spatial resolution, since the surface tension scales as force per length unit. That, however, would require significantly more lattice sites in each direction, thus enormously increasing the computational effort.

Furthermore, a number of parameters have a strong impact on the propagation of the magnetocapillary swimmer. First, it is the particle radius $R$ which should be larger than the thickness of the diffuse interface ($\approx 5\Delta x$)~ \- \cite{JaHa11,GJFH13} and large enough to provide a spatial resolution required to reproduce the correct surrounding flow field.
At the same time, $R$ has to be small enough to assure a comparison with the experimental situation, where the radius is small compared to the system size.
Second, in view of simulations of long-ranged capillary phenomena the total size of the simulated fluid or the box size is very crucial. Using periodic boundary conditions in lateral directions of the box (Fig.~\ref{fig_1} a), the box side length should be large to ensure saturation of the interface from the point of contact with the particles towards the edges. And although the LB-method is nicely scalable with the box size, very large system sizes require enormous computational times. The third parameter that should be carefully chosen is the $Bo$-number, which can be tuned either by the particle density or by its radius. For a better interface resolution a notably curved interface profile is desired, hence, large $Bo$-number, while exceeding $Bo^{\mathrm{crit}}\approx 0.21$ leads to sinking of particles. Taking into account the listed criteria and the available computational resources is the base for our choice of parameters as given in the end of sect. 2. 

\section{Conclusions and Outlook}
We demonstrated that our LB simulations are capable of reproducing the rich dynamics of magnetocapillary microswimmers by taking into account all relevant physical ingredients. We proved in particular that the existence of small ferromagnetic contributions in the particle properties ($\vec{\mu}\ind{int}\neq 0$) captures the characteristic swimmer reorientations observed experimentally~\cite{GrLa15}. Moreover, we claim that when the magnetization of the beads is only induced by an external magnetic field ($\vec{\mu}\ind{ind}\neq 0$, $\vec{\mu}\ind{int}=0$), the swimmer is also capable of swimming and its motion is then characterised
by the maximum swimmer velocity to be centered around the particle’s inverse coasting time in the range of higher driving frequencies. For lower driving frequencies and a high ratio of surface tension to magnetic forces, the swimmer motion is determined by the strength of the magnetocapillary particle interactions.

As an outlook, yet another regime of motion might be numerically studied. In that case, an additional small static magnetic field is applied along the z-axis (Fig.~\ref{fig_1}) leading to sizable \textit{individual} rotations of each bead in the plane of the interface. In this setup a swift swimmer motion is reached experimentally presumably because of strong hydrodynamic flows~\cite{GrHu17}.

\section*{Acknowledgements}
This work was financially supported by the DFG Priority Programme SPP 1726 ``Microswimmers—From Single Particle Motion to Collective Behaviour'' (HA 4382/5-1). We further acknowledge the Jülich Supercomputing Centre (JSC) and the High Performance Computing Centre Stuttgart (HLRS) for the allocation of computing time.

\section*{Author contribution statement}
All the authors were involved in the preparation of the manuscript. All the authors have read and approved the final manuscript.

\section*{Appendix A: Calculation of spring constants}
\renewcommand{\theequation}{A.\arabic{equation}}
\setcounter{equation}{0}
\begin{figure}[htb]
\centering
\includegraphics[width=0.48\textwidth]{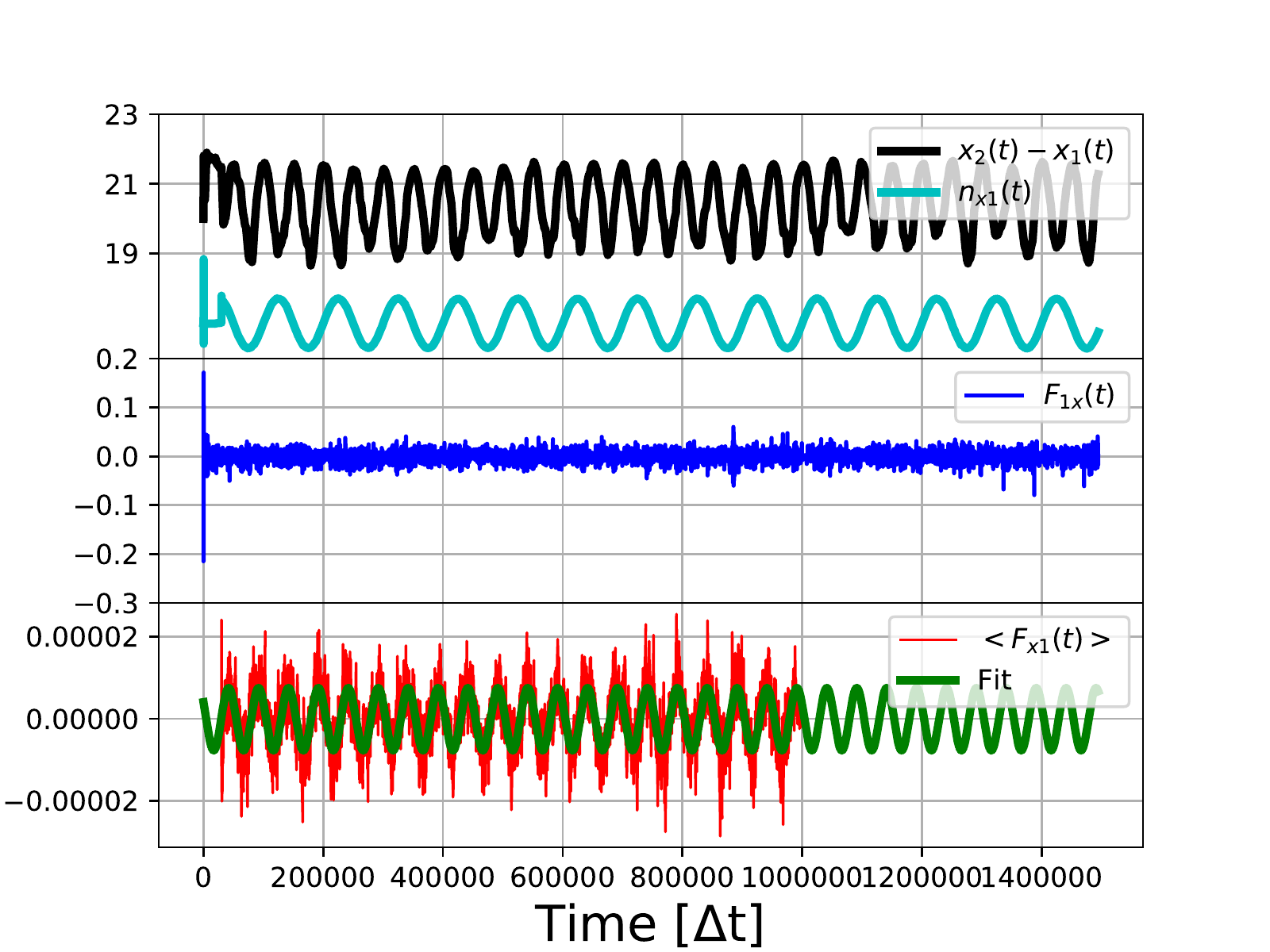}
\caption{Extraction of spring constants characteristic for the motion of particles 1 and 2 within the swimmer. Upper panel: mutual displacement $x_2(t)-x_1(t)$ together with the director vector $n_{x 1}(t)$ showing its propagation driven by $B(t)$. Middle panel: non-averaged force $F_{x 1}(t)$. Low panel: Averaged and fitted forces acting on particle 1. LB-parameters: $Bo=0.16$, $L=2.1\times (2R)$, $|B(t)|/|B|=0.36$, $T=100000$~$\Delta t$.}
\label{fig_1_AA}
\end{figure} 

For the spring potential between \textit{e.g.} particles 1 and 2 (Fig. \ref{fig_1}) defined as
\begin{equation}
 \displaystyle \phi(\vec{r}_1-\vec{r}_2)=\phi_{12}=\frac{1}{2}k \left(|\vec{r}_1-\vec{r}_2|-L\right)^2,  
\label{eq_1_AA}
\end{equation}
with $L$ being the equilibrium distance between the centres of particles, we define the force acting on particle 1 via
\begin{equation}
 \displaystyle \vec{F}_1(t)= - \vec{\nabla}_1 \phi(\vec{r}_1-\vec{r}_2).  
\label{eq_2_AA}
\end{equation}
In general, spring constants have different components $k_x$, $k_y$ and $k_z$, so that 
\begin{equation}
\begin{split}
 \displaystyle \vec{F}_1(t)= - & \left(1-\frac{L}{|\vec{r}_1(t)-\vec{r}_2(t)|}\right) \times \\ 
& \begin{pmatrix}
 k_{1x} & 0 & 0 \\
 0 & k_{1y} & 0 \\
 0 & 0 & k_{1z}
\end{pmatrix} 
\begin{pmatrix}
x_1(t)-x_2(t) \\
y_1(t)-y_2(t) \\
z_1(t)-z_2(t)
\end{pmatrix}.
\label{eq_2a_AA}
\end{split}
\end{equation}
%\SZ{Just as a remark: Apparently, the equilibrium length L is independent under rotation of the line connecting both particles, but the spring constants are not. I know that one component of $k$ is associated to the surface, but for this component the equilibrium length probably does not relate to L.}
The x-component is defined according to eq. (\ref{eq_2_AA}) as
\begin{equation}
\displaystyle k_{1x} = \frac{-F_{1x}(t)}{\left(1-\frac{L}{|\vec{r}_1(t)-\vec{r}_2(t)|}\right)\left(x_1(t)-x_2(t)\right)}.
\label{eq_3_AA}
\end{equation}
The time average is defined using
\begin{equation}
\displaystyle \langle k_{1x}\rangle = \frac{1}{t_{\infty}-t_0}\int_{t_0}^{t_{\infty}}\frac{-F_{1x}(t)dt}{\left(1-\frac{L}{|\vec{r}_1(t)-\vec{r}_2(t)|}\right)\left(x_1(t)-x_2(t)\right)}.
\label{eq_4_AA}
\end{equation}
Using expression (\ref{eq_4_AA}) one can determine \textit{e.g.} $\langle k_{12 x}\rangle$ between particles 1 and 2 from their trajectories upon the swimmer motion. The unit for the $k$-constant in LB-simu\-la\-tions is $[\rho_0 \frac{\Delta x^3}{3\Delta t^2}]$.

Fig.~\ref{fig_1_AA} demonstrates how the effective spring constant $k_{12 x}$ related to the interaction between particles 1 and 2 can be calculated. For this the relative displacement between particles 1 and 2 $x_2(t)-x_1(t)$ is steadily measured while the swimmer moves (Fig.~\ref{fig_1_AA}, upper panel). At the same time the total force acting on particle 1 is recorded (Fig.~\ref{fig_1_AA}, middle panel) and averaged over the period of the external magnetic field $2\pi/\omega$. Inserting the obtained expressions for the force and the mutual displacements into eq.~(\ref{eq_4_AA}), we obtain the values of $k_{12 x}(L)$ as a function of the swimmer size. Since the capillary potential gets very distorted at low swimmer sizes, the expressions of total averaged forces $\langle F_{x 1}(t)\rangle$ are very noisy~\cite{KHV10}. For moderate and large swimmer sizes ($L\approx 2\times (2R)$) the picture is represented by Fig.~\ref{fig_10} and yields values of the order $k_{12 x}\approx 10^{-5}$ l.u. 
%\AS{Discussion on how exact the represented procedure is (total force, 2 vs 3 particles ...)} \\

%\MH{What is F1(t)? The total force? the MC force? the magnetic one? Also, how is the value of F1 obtained from the simulations? }

\section*{Appendix B: Implementation of magnetic for\-ces in case of a finite internal magnetic moment}
\renewcommand{\theequation}{B.\arabic{equation}}
\setcounter{equation}{0}
In the experiments on magneticapillary swimmers~\cite{GrLa15}, there are indications of the existence of a small permanent magnetic moment, such that the total magnetic moment of each particle reads
\begin{equation}
\displaystyle \vec{\mu}\ind{tot} = {\color{black}\vec{\mu}\ind{ind}} + {\color{black}\vec{\mu}\ind{int}},
\label{eq_1_AB}
\end{equation} 
whereby the induced magnetic moment is the result of the external field, i.e $\vec{\mu}\ind{ind} \sim \vec{B}$, while the internal magnetic moment $\mu\ind{int}\ll \mu\ind{ind}$ is not a function of the external field and is always present.

Thus, the force exerted by the moment $\vec{\mu}\ind{tot}_j$ on the magnetic moment $\vec{\mu}\ind{tot}_i$ is
\begin{equation}
\displaystyle \vec{F}_{ji}^{\mathrm{magn. tot}} = -\vec{\nabla} \left(-({\color{black}\vec{\mu}\ind{ind}_i}+{\color{black}\vec{\mu}\ind{int}_i}) \cdot \vec{B}\ind{tot}_j\right), 
\label{eq_2_AB}
\end{equation}
where 
\begin{equation}
\begin{split}
\displaystyle \vec{B}\ind{tot}_j/\left(\frac{\mu_0}{4\pi|\vec{r}|^3_{ij}}\right) = 
& \left[3\vec{e}_{ij}({\color{black}\vec{\mu}\ind{ind}_j}\cdot \vec{e}_{ij})-{\color{black}\vec{\mu}\ind{ind}_j}\right]\\ 
& +\left[3\vec{e}_{ij}({\color{black}\vec{\mu}\ind{int}_j}\cdot \vec{e}_{ij})-{\color{black}\vec{\mu}\ind{int}_j}\right]. 
\end{split}
\label{eq_3_AB}
\end{equation}
The resulting force exerted by the moment $\vec{\mu}\ind{tot}_j$ on the magnetic moment $\vec{\mu}\ind{tot}_i$ is 
\begin{widetext}
\begin{equation}
%\footnotesize
\begin{split}
\displaystyle & \vec{F}_{ji}^{\mathrm{magn. tot}}/ \left(\frac{3\mu_0}{4\pi |\vec{r}_{ji}|^4}\right) = \\ 
& \left({\color{black}\vec{\mu}\ind{ind}_i} \left({\color{black} \vec{\mu}\ind{ind}_j} \cdot \vec{e}_{ji} \right) + {\color{black}\vec{\mu}\ind{ind}_j} \left({\color{black}\vec{\mu}\ind{ind}_i} \cdot \vec{e}_{ji} \right) - 5 \vec{e}_{ji} \left({\color{black}\vec{\mu}\ind{ind}_j} \cdot \vec{e}_{ji} \right) \left({\color{black}\vec{\mu}\ind{ind}_i} \cdot \vec{e}_{ji} \right)+\vec{e}_{ji}({\color{black}\vec{\mu}\ind{ind}_i} \cdot {\color{black}\vec{\mu}\ind{ind}_j})\right)\\
& + \\
& \left( {\color{black}\vec{\mu}\ind{ind}_i} \left({\color{black}\vec{\mu}\ind{int}_j} \cdot \vec{e}_{ji} \right) + {\color{black}\vec{\mu}\ind{int}_j} \left({\color{black}\vec{\mu}\ind{ind}_i} \cdot \vec{e}_{ji} \right) - 5 \vec{e}_{ji} \left({\color{black}\vec{\mu}\ind{int}_j} \cdot \vec{e}_{ji} \right) \left({\color{black}\vec{\mu}\ind{ind}_i} \cdot \vec{e}_{ji}\right)+\vec{e}_{ji}({\color{black}\vec{\mu}\ind{ind}_i} \cdot {\color{black}\vec{\mu}\ind{int}_j})\right)\\
& + \\
& \left( {\color{black}\vec{\mu}\ind{int}_i} \left({\color{black}\vec{\mu}\ind{ind}_j} \cdot \vec{e}_{ji} \right) + {\color{black}\vec{\mu}\ind{ind}_j} \left({\color{black}\vec{\mu}\ind{int}_i} \cdot \vec{e}_{ji} \right) - 5 \vec{e}_{ji} \left({\color{black}\vec{\mu}\ind{ind}_j} \cdot \vec{e}_{ji} \right) \left({\color{black}\vec{\mu}\ind{int}_i} \cdot \vec{e}_{ji}\right)+\vec{e}_{ji}({\color{black}\vec{\mu}\ind{int}_i} \cdot {\color{black}\vec{\mu}\ind{ind}_j})\right)\\
& + \\
& \left( {\color{black}\vec{\mu}\ind{int}_i} \left({\color{black}\vec{\mu}\ind{int}_j} \cdot \vec{e}_{ji} \right) + {\color{black}\vec{\mu}\ind{int}_j} \left({\color{black}\vec{\mu}\ind{int}_i} \cdot \vec{e}_{ji} \right) - 5 \vec{e}_{ji} \left({\color{black}\vec{\mu}\ind{int}_j} \cdot \vec{e}_{ji} \right) \left({\color{black}\vec{\mu}\ind{int}_i} \cdot \vec{e}_{ji}\right)+\vec{e}_{ji}({\color{black}\vec{\mu}\ind{int}_i} \cdot {\color{black}\vec{\mu}\ind{int}_j})\right).
\end{split}
\label{eq_4_AB}
\end{equation}
\end{widetext}
Similarly, magnetic torques should read
\begin{equation}
\displaystyle \vec{T}_{ji}^{\mathrm{magn.tot}} = \left[ \left({\color{black}\vec{\mu}\ind{ind}_i}+{\color{black}\vec{\mu}\ind{int}_i}\right) \times \left(\vec{B}\ind{ind}_j+\vec{B}\ind{int}_j+\vec{B}\right)\right].
\label{eq_5_AB}
\end{equation}
The resulting total magnetic torque is
\begin{equation}
\begin{split}
\displaystyle & \vec{T}_{ji}^{\mathrm{magn. tot}}/ \left(\frac{\mu_0}{4\pi |\vec{r}_{ji}|^3}\right) = \\ 
& \left(3\left({\color{black}\vec{\mu}\ind{ind}_j} \cdot \vec{e}_{ji}\right)\left[{\color{black}\vec{\mu}\ind{ind}_i}\times \vec{e}_{ji} \right] - \left[{\color{black}\vec{\mu}\ind{ind}_i} \times {\color{black}\vec{\mu}\ind{ind}_j}\right]\right)+\\
&\left(3\left({\color{black}\vec{\mu}\ind{int}_j} \cdot \vec{e}_{ji}\right)\left[{\color{black}\vec{\mu}\ind{ind}_i}\times \vec{e}_{ji} \right] - \left[{\color{black}\vec{\mu}\ind{ind}_i} \times {\color{black}\vec{\mu}\ind{int}_j}\right]\right)+\\
&\left(3\left({\color{black}\vec{\mu}\ind{ind}_j} \cdot \vec{e}_{ji}\right)\left[{\color{black}\vec{\mu}\ind{int}_i}\times \vec{e}_{ji} \right] - \left[{\color{black}\vec{\mu}\ind{int}_i} \times {\color{black}\vec{\mu}\ind{ind}_j}\right]\right)+\\
&\left(3\left({\color{black}\vec{\mu}\ind{int}_j} \cdot \vec{e}_{ji}\right)\left[{\color{black}\vec{\mu}\ind{int}_i}\times \vec{e}_{ji} \right] - \left[{\color{black}\vec{\mu}\ind{int}_i} \times {\color{black}\vec{\mu}\ind{int}_j}\right]\right)+\\
&\left[\left({\color{black}\vec{\mu}\ind{ind}_i}+{\color{black}\vec{\mu}\ind{int}_i}\right)\times \vec{B}\right]/\left(\frac{\mu_0}{4\pi |\vec{r}_{ji}|^3}\right).
\end{split}
\end{equation}

\printbibliography

\end{document}